\documentclass[preprint,12pt,authoryear]{elsarticle}
\usepackage[T1]{fontenc}
\usepackage[latin9]{inputenc}
\usepackage{amstext}
\usepackage{graphicx}
\usepackage{esint}
\usepackage{pdflscape}

\makeatletter
\usepackage[latin9]{inputenc}

\usepackage{subfig}

\journal{}

\begin{document}
\begin{frontmatter}

\title{Analysis of mesoscale forecasts using ensemble methods}

\author{Markus Gross}
\address{CICESE (Centro de Investigación Científica y de Educación Superior de Ensenada), Physical Oceanography, Carretera Ensenada-Tijuana 3918, Ensenada BC 22860, MEXICO}
\begin{abstract}
Mesoscale forecasts are now routinely performed as elements of operational forecasts and their outputs do appear convincing. However, despite their realistic appearance at times the comparison to observations is less favorable. At the grid scale these forecasts often do not compare well with observations.  This is partly due to the caotic system underlying the weather. Another key problem is that it is impossible to evaluate the risk of making decisions based on these forecasts because they do not provide a measure of confidence. Ensembles provide this information in the ensemble spread and quartiles. However, running \emph{global} ensembles at the meso or sub mesoscale involves substantial computational resources. National centers do run such ensembles, but the subject of this publication is a method which requires significantly less computation. The \emph{ensemble enhanced mesoscale system} presented here aims \emph{not} at the creation of an improved mesoscale forecast model. Also it is \emph{not} to create an improved ensemble system. Here an ensemble is created from one mesoscale forecast with the aim of interrogating the probabilities of the forecast. The diagnostics developed in this publication is the generation of the confidence intervals via cumulative probability density functions (pdf), detection of extrema and selective ensembles. The subject of this publication is the analysis of those diagnostics, their dependence on the domain size of the ensemble and the number of ensemble members. The analysis strategies introduced are termed: \emph{subgrid ensemble enhanced mesoscale forecast}, \emph{extreme value analysis of the ensemble pdf} and \emph{observation constrained ensemble forecast}. 
\end{abstract}

\end{frontmatter}

\section{Introduction}

Mesoscale forecast provide a stunning amount of detail and realism.
Less dependence on parametrizations of physical processes and more resolved physics leads
to a representation of weather features that appear convincing. When
compared to observation, however, it is often found that locality, timing
and intensity are at times not as accurate as may be desirable. As
noted by \citet{ASL:ASL72}, explicit treatment of convection in forecasts
does not necessarily provide more accurate point forecast, but rather
a more accurate depiction of the physics. Also, with sufficient lead
time the single deterministic solution may vary from
the observed weather to the extend that even climate means often produce
smaller errors (cf \citet{doi:10.1175/1520-0493(1974)102<0409:TSOMCF>2.0.CO;2,TELLUSA10143}).
This then raises the question, how reliable is this forecast? A question
a single deterministic forecast cannot answer. However, sometimes
it is important to know how reliable the forecast is. For example,
during a week in January 2016 severe rainfall was predicted for the
region of Ensenada. 
Some of the available deterministic models at the time predicted accumulated
precipitation of $70$mm or more over one or two days. This created
anxiety amongst the vulnerable public. Also emergency responders were
communicating these high values, demonstrating the high level of confidence
attributed to these forecasts. The anxiety was increased in particular because the events occurred during the peak
of a very strong el Ni\~{n}o event, following months of alarming speculation about the
potential impacts of expected severe precipitation events commonly
associated with el Ni\~{n}o in this region \citep{ASL2:ASL2656}. When compared with
available ensemble prediction systems, \citet{doi:10.1175/2008MWR2682.1},
it could be seen that these $70$mm were likely a severe over
estimate of a single deterministic forecast. For risk assessment and
planning, knowledge of the probabilities associated with forecast events is essential (\citet{QJ:QJ200212858101,MET:MET140}). Deterministic models cannot provide these due to their deterministic nature. Attempts are often made to compute probabilities of precipitation in the proximity by sampling the deterministic forecast. This however still assumes that the one forecast is an accurate representation of the weather at that time, which is not necessarily the case due to the chaotic nature of the weather system. An alternative to the deterministic model is an ensemble system. However, the current global ensemble systems are running at $\approx 20$km grid resolution with an effective resolution of $60-80$km. Therefore, whilst producing some confidence in their forecasts and a level of certainty, the spacial uncertainty can still lead to large local forecast errors. Downscaling from
the perturbed \emph{global} model, \citet{doi:10.1175/WAF-D-15-0102.1},
creates a significant amount of computational work, even at coarse
target resolutions. This motivates the creation of a system which
starts at the mesoscale, perturbs the solution (the initial condition for the ensemble) and evaluates the ensemble output.
This has the advantage that it can increase the resolution and at
the same time provide insight on the reliability of the forecast provided
by the deterministic model. It can be reasonably expected that the
high resolution nests spin up the features missing in the coarser
grids in a few hours (\citet{doi:10.1175/MWR2830.1}), providing
 useful forecast data for the remainder of the forecast period. Starting
from the mesoscale creates an additional parameter which
potentially has a detrimental impact on the output: domain size. If
the domain size is small, the unperturbed boundary condition will constrain
the solution and limit the statistical value of the experiment and
create overly confident predictions (\citet{QJ:QJ2238}). As stated
above: The aim of this work is not to generate the ideal convection
resolving ensemble (cf \citet{QJ:QJ2238,doi:10.1175/MWR-D-12-00031.1}),
which inevitably needs data from global ensembles. The \textit{ensemble enhanced mesoscale forecast } proposed here can be desirable in
at least three scenarios: 1) where no sufficient resource exists to run
the required \emph{global} ensemble (or downscale each member of the \emph{global}
ensemble to the required local resolution), 2) when time is of the essence (for example in emergency response situations or forecaster-requested
on-demand modeling support for the weather desk), or 3) in the case of
remote embedded forecast units which rely on satellite transmission
of the initial conditions.

Therefore the underlying assumption for this work is that in some situations there is not sufficient time or computing resource available to compute a high resolution convection resolving limited area model ensemble starting from global initial data. Therefore the starting point is a single deterministic mesoscale forecast. An ensemble system is then used in order to evaluate the uncertainty of this forecast without degrading the original forecasts resolution, as a neighborhood forecasting method (\cite{MET:MET200512308}) would. Ideally the ensemble system will provide a forecast with even higher resolution as the original deterministic one, as a key aim is the forecast at a \emph{specific location} and not area and especially not area of tens or even hundreds of kilometers squared. This high resolution, combined with the probabilistic guidance based on the intrinsic chaotic nature of the system (and not just some geographic evaluation) will enable the use of the forecast for emergency response. The novel aspects of this work are the introduction of what can be termed a subgrid ensemble and the further processing of this ensemble output using available observations to constrain the ensemble members and the extreme value analysis of the pdf. Here the effect of domain size, ensemble members and lead time will be illustrated, with the aim to produce \emph{point} forecasts from a \emph{minimal} amount of data and a \emph{minimum} amount of computational work. Furthermore, the probabilistic guidance generated is based on the evaluation of the intrinsic chaotic nature of the weather system, which is not available in the traditional neighborhood approaches (cf. \cite{MET:MET200512308} and \cite{doi:10.1175/2007MWR2123.1}, for example).

In the following the methodology is presented, including the study site. This is followed by results which illustrate the methodology and demonstrate how the forecast is improved. This is followed by brief conclusions.

\section{Methodology}
\begin{figure}
\centerline{\includegraphics[width=\textwidth]{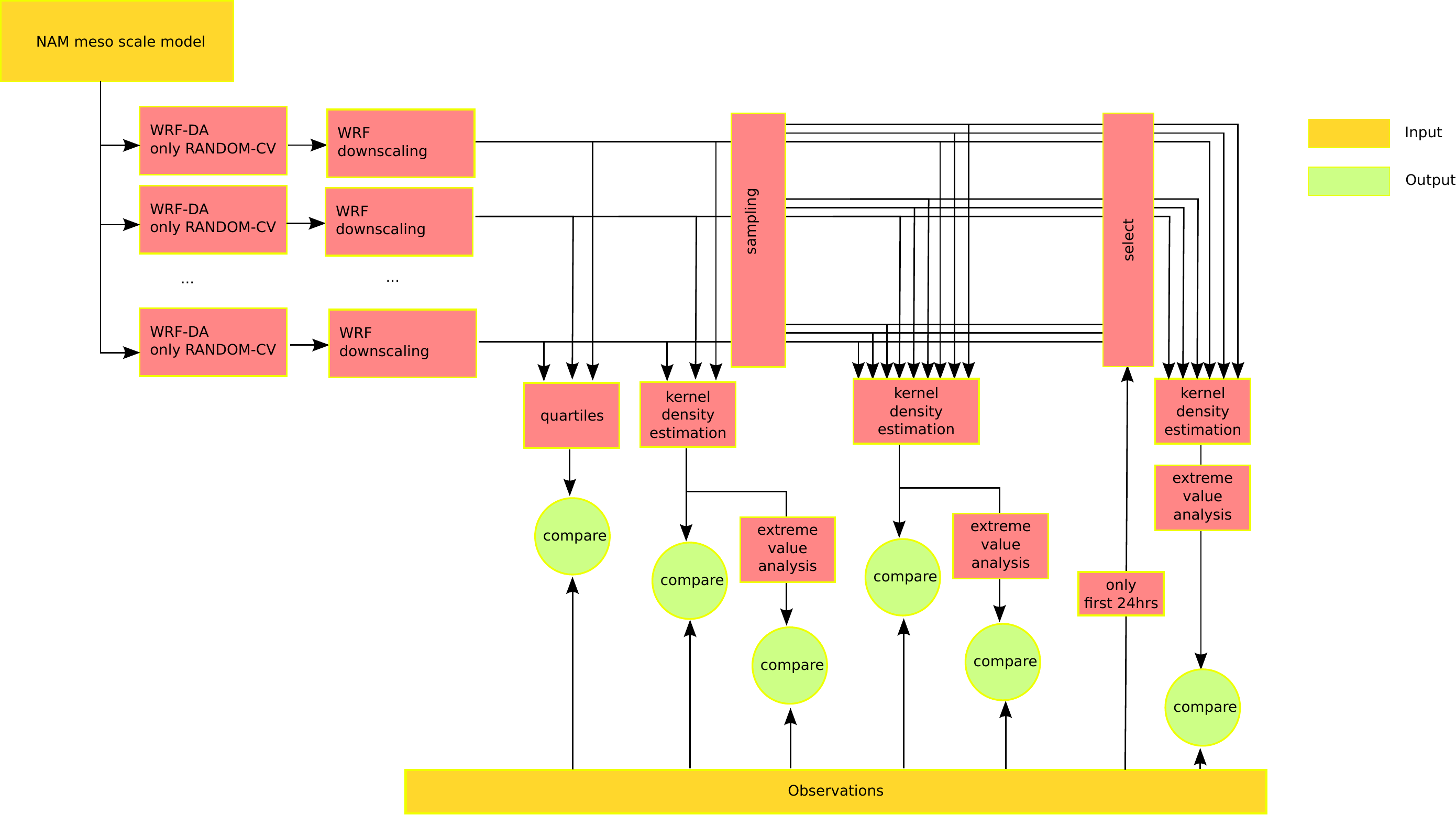}}
\caption{Methodology\label{fig:method}}
\end{figure}

The methodology is illustrated in Figure \ref{fig:method}.
A single deterministic forecast has been perturbed by adding random noise to the analysis in control variable space to generate initial conditions for each ensemble member. These initial conditions are then downscaled to increase the resolution and later provide grid points for sampling without surpassing the resolution of the initial data in an approach similar to neighborhood forecasting, however without degrading the resolution of forecast beyond the resolution of the initial forecast. The output is then analyzed with regard to initial domain size and number of ensemble members. Initially, the quartiles and kernel density estimation is run with an increasing number of ensemble members without any special selection procedure. A local extreme value analysis is performed and shown to indicate potential outcomes (branches) in the ensemble. Then members are selected according to the comparison against the first 24 hours of observations in order to improve the forecast for forecast hours 25 to 60.

\subsection{The study area}

The case studies have been performed for the Port of Ensenada, Baja California, Mexico. The site was chosen due to the importance of precipitation forecasts in this area. Precipitation is usually sparse and, when it does occur, events are intense. Preparation and planning are important both for agriculture
and - during extreme  events - for the emergency services. The combination of the coastal setting with mountainous and desert regions in close proximity challenges the forecast models.

\subsection{Model setup}

\begin{figure}
\subfloat[Largest domain (n=160)]{\protect\centering{}\protect\includegraphics[width=0.5\textwidth]{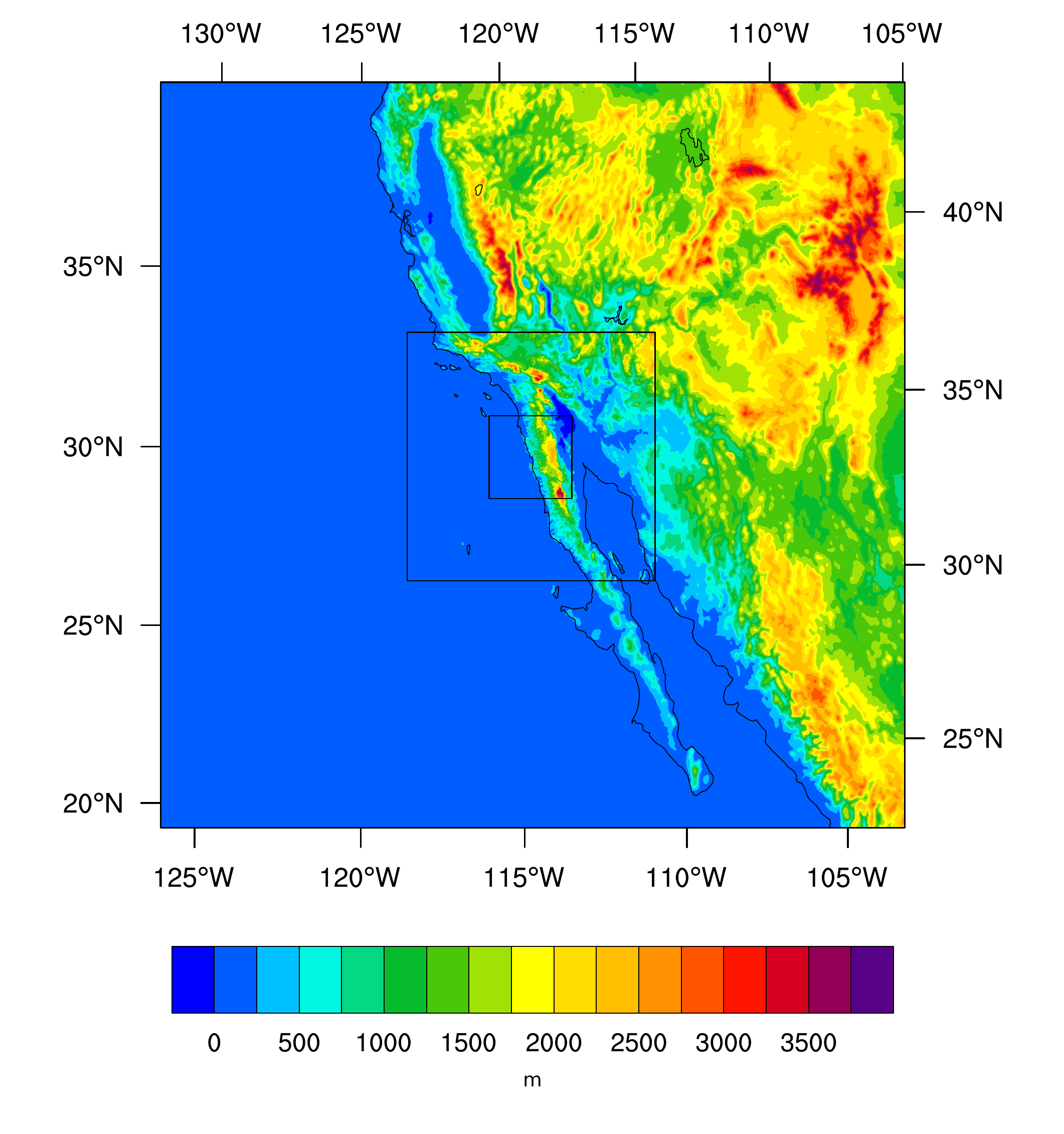}\protect}
\subfloat[Smallest domain (n=40)]{\protect\centering{}\protect\includegraphics[width=0.5\textwidth]{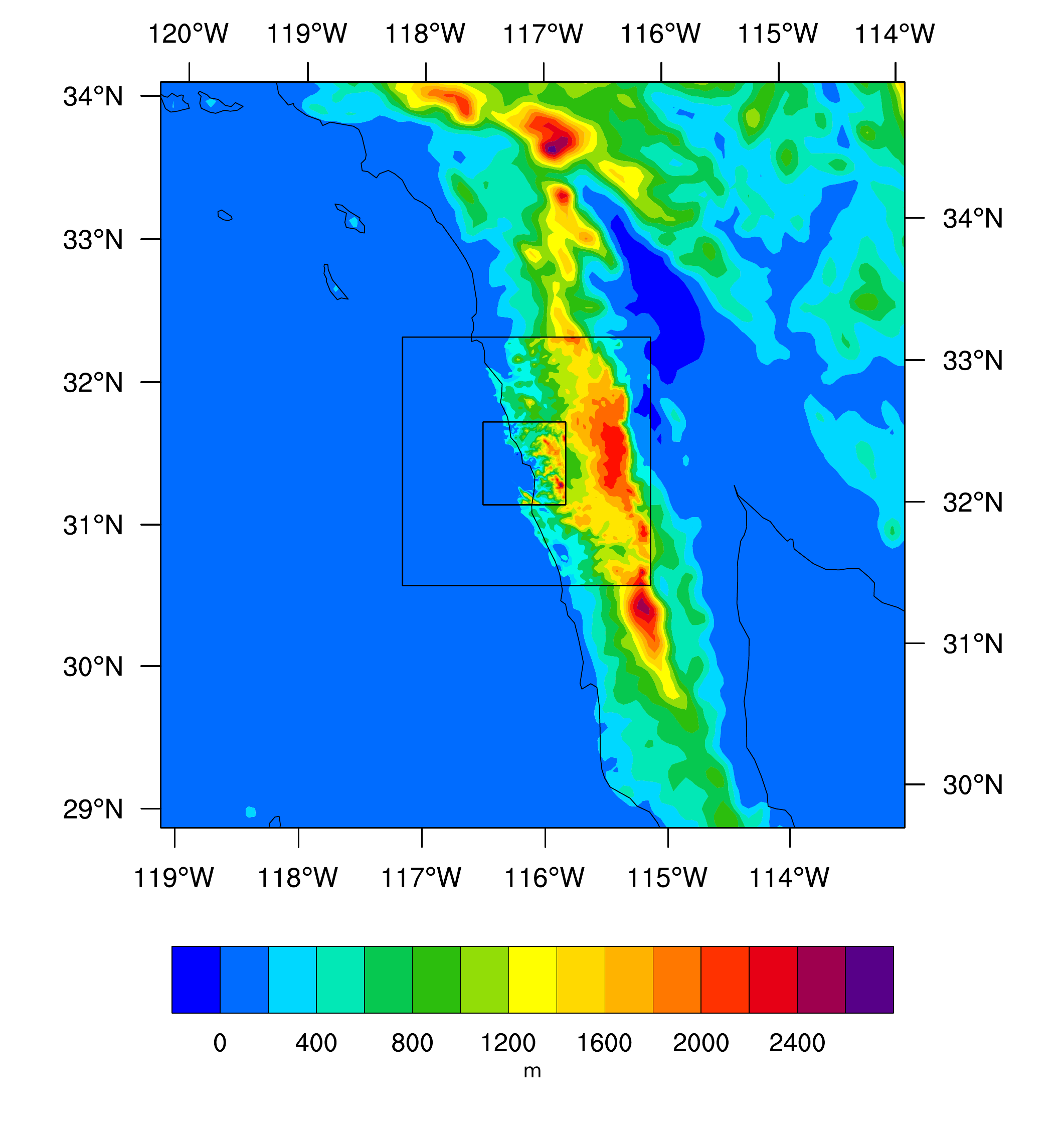}\protect}\protect\caption{Model domains showing orography of the study area.\label{fig:Model-domains} The two nests (d02 and d03) are indicated by black lines. All domains have the same resolution for the parent (d01) and respective nests.}
\end{figure}

The Weather Research and Forecasting (WRF) model \citep{Skamarock08} version
3.5.1 was run over the study area with three nests in two-way nesting.
The resolutions of the domains and general model configuration is listed in Table~\ref{tab:config}. Six different domain sizes were run, $n=40,$ $80$, $100$,
$120$, $140$ and $160$. This referes to a grid with $3\times n$ grid points West to East and North to South, $9n^2$ grid points in total. The grid is centered over
Ensenada, reference latitude of $31.9394$ degrees and reference longitude
of $-116.5863$ degrees.  The domains,
largest and smallest, are shown in Figure \ref{fig:Model-domains}.
 The initial data and boundary conditions
were taken from the 5km resolution North American Mesoscale (NAM) model CONUS Nest, run by the
US National Weather Service - NCEP (WMC). The boundary update interval is one hour for the first $36$ hours and then three hours until forecast hour 60.  WRFDA \citep{Barker04,Huang09}, WRF Data Assimilation, 
was used to create the perturbed ensemble members, with only a minor
modification to the random number generator seed, which was modified
to create a seed of full length.
No observations were assimilated into the downscaling runs. The initial condition is provided by a Numerical Weather Prediction model which already assimilates observations.

\begin{table}[h]
\begin{tabular}{@{}lccc@{}}
\hline 
 & d01 & d02 & d03\\
\hline
resolution & $5000$m & $1666.66$m & $555$m \\
time step & $20$s & $20/3$s & $20/9$s\\
micro physics & WSM 3-class simple ice & same as d01 & same as d01\\
ra\_lw\_physics & rrtm  & same as d01 & same as d01 \\
ra\_sw\_physics & Dudhia & same as d01 & same as d01\\
time between radiation & $1$min & same as d01 & same as d01\\
surface-layer & MM5 Monin-Obukhov & same as d01 & same as d01\\
land-surface & thermal diffusion & same as d01 & same as d01\\
boundary-layer & YSU & same as d01 & same as d01\\
cumulus & Kain-Fritsch (new Eta) & same as d01 & no cumulus\\
non-hydrostatic & .true. & .true. & .true.\\ 
dfi\_opt & twice DFI (TDFI) &  & \\
\hline
\end{tabular}
\caption{Model setup for the three domains d01, d02 and d03.\label{tab:config}}{}
\end{table}

\subsection{Analysis of the model output}
The ensemble results were then analyzed with respect to accumulated precipitation. The ensemble mean, minimum
and maximum, as well as the quartiles, probability density function
(pdf) and cumulative distribution function (cdf) were compared to observations. The observations were obtained from the national meteorological service (Servicio Meteorologico Nacional), station name: P.LOPEZ ZAMORA, Ensenada, BC, Mexico, longitude: $116\protect\textdegree36'12"$,
latitude: $31\protect\textdegree53'29"$, altitude: $32$m. In all Figures presenting forecast data these observations are plotted in blue.

\subsubsection{Quartiles}
Quartiles are defined by the quartile function:
\begin{equation}
q\left(x_{q}\right)=p_{acc}\left(x=x_{q}\right)\label{eq:trad}
\end{equation}
where $q\left(x_{q}\right)$ is the accumulated precipitation with
probability of $x_{q}$ and $p_{acc}\left(x_{i}\right)$ is the sorted
array of accumulated precipitation values from the ensemble members
at discrete probabilities $x_{i}$, where the spacing and value of
the $x_{i}$ depends on the ensemble size. $p_{acc}\left(x=x_{q}\right)$
is then evaluated using linear interpolation.
\subsubsection{Pdf and cdf}
The distribution of precipitation is truncated at zero as it cannot be negative. Therefore a truncated kernel
density estimation (kde) is performed, with lower bound zero. A normal
Gaussian kernel is assumed, such that the pdf $\phi$ is 
\begin{equation}
\phi=\frac{1}{\sqrt{2\pi}}\exp^{-\frac{x^{2}}{2}}.
\end{equation}
Linear combination correction (\citet{Jones03}) has been performed
to represent the truncated distribution, $\phi_{t}$, which has a
lower bound of zero. The obtained pdf has the advantage that it can
be readily integrated into the cdf
$\hat{\phi}$ 
\begin{equation}
\hat{\phi}\left(x\right)=\int_{0}^{x}\phi_{t}\left(u\right)du\label{eq:cpd}
\end{equation}
and more accurate quartiles can be obtained by evaluation of this integral.
\subsubsection{Subgrid ensemble enhanced mesoscale forecast}
In the sub-grid ensemble the  ensemble size is extended. Not by adding more ensemble members, but by allowing
for a geographic uncertainty. The ensemble members from neighboring grid points are combined to form a larger ensemble. 

Since the ensemble downscales the mesoscale forecast the solution can be sampled without reducing
the resolution beyond the original mesoscale forecast. This sampling is similar to the neighborhood forecasting approach of \cite{MET:MET200512308}. However, here the resolution is at least the resolution of the driving mesoscale model, not less. Furthermore it is important to note that here the chaotic nature of the weather system is sampled by the ensemble, which is not the case in the traditional  neighborhood forecasting approach, nor in any other method which attempts to extract probabilities from deterministic data. The third domain can be sampled in $\left(3\times 3\right)^2=81$ points before the resolution of the original mesoscale forecast is reached.
This is the \emph{subgrid ensemble enhanced mesoscale forecast}.

\subsubsection{Extreme value analysis of the ensemble pdf}
 The pdf obtained from the ensemble can be further analyzed for local extrema (maxima). These extrema will be shown to carry useful information, highlighting potential outcomes which would be lost if only the quartiles and ensemble mean would be considered.

\subsubsection{Constrained ensemble}
Finally the data is analyzed taking observations into account. Ensemble members
are selected according to their difference to observation. A ranking,
$s_{e}$, is computed for each ensemble member $e$ over the first $24$
samples (here one per hour) 
\[
s_{e}=\sum_{i=1}^{24}s_{e,i},
\]
where 
\begin{equation}
s_{e,i}=\left\{ \begin{array}{ccc}
1 & if & obs\left(t_{i}\right)-\sigma<p_{acc}\left(e,t_{i}\right)<obs\left(t_{i}\right)+\sigma\\
0 & \textrm{otherwise}
\end{array}\right.\label{eq:corr}
\end{equation}
with $obs$, $t$, $i$, $\sigma$, $p_{acc}\left(e,t_{i}\right)$
denoting the observed accumulated precipitation, discrete time and
its index $i$, standard deviation of all ensemble members at the
respective time step and accumulated precipitation of ensemble member
$e$ at time $t_{i}$, respectively. If $s_{e}>$$s_{thresh}$, where
$s_{thresh}$ is a threshold to be defined, the member is retained
in the ensemble. The threshold is a free parameter, as illustrated in the results below. The aim of this final exercise it to improve the
 the forecast at a time when observations are available.
Naturally, the error or uncertainty is largest with longer lead time 
and some improvement becomes necessary. It will be shown that this can be achieved using this selection process.

\section{Results and Discussion}

First, a comparison of model versus observations under varying domain and ensemble sizes
for two key dates, initial time 4th of January 00:00 UTC and initial time of 7th of January 00:00 UTC are presented.

 Then the benefit of the \emph{subgrid ensemble enhanced mesoscale forecast} is shown.
This improves the ensemble, meaning that more outcomes resemble the observations. Line plots of the time series of the ensemble members suggest some clustering, i.e. some outcomes appear more likely than others.

Following this observation the local maxima in the pdf are analyzed, the \emph{extreme value analysis of the ensemble pdf}. This reveals several possible outcomes and their bifurcation points.

Finally,  in order to improve the later part of the forecast, 
observations made during  the first 24 hours are ``assimilated'' into the geographically
extended ensemble and results of this exercise are also presented.

\subsection{Point forecasts\label{sub:Point-forecasts}}

Precipitation is amongst the most difficult variables to predict and yet arguably one of
the most important aspects of numerical weather forecasting. This difficulty is often glossed
over by reporting the probability of precipitation (PoP; \citet{WOM15}),
which is computed by evaluating the precipitation over a larger geographic
area and time window (and is associated with great difficulty in its communicating to
and understanding by the general public (\citet{doi:10.1175/2008BAMS2509.1})). However,
this has only limited success and, more critically, significantly
less value for the individual who is mostly interested in precipitation
in very specific areas and less concerned about precipitation elsewhere.
For this reason, the interest here lies with forecasts at specific
locations, which can then also be readily compared to ground-based
observations (it is less clear how PoP can be compared to observations
other than satellite observations with their larger margin of error).
Here the focus is on accumulated precipitation, which simplifies the
comparison as it relaxes the timing component, the ``Double Penalty
Issue'' \citep{doi:10.1175/1520-0477(1998)079<0253:RRTNWP>2.0.CO;2}:
\begin{itemize}
\item The precipitation event may \emph{not appear} at exactly the same time as in the observation and then
\item  \emph{it appears} in the model (minutes earlier or later) when it
does not appear in the observational record.
\end{itemize}
This complicates objective verification of instantaneous precipitation significantly.

\newcommand{\custcap}[2]{Forecast with initial time: #2 Jan 2016 00:00 UTC. Domain size of $n=#1$, domain d03. From a) to d) the number of ensemble members considered increases, from 6 to 11, then 16 and finally 20. The observed precipitation is plotted in blue, the cumulative density function interval from 0.25 until 0.75 in transparent red fill, 0.25 to 0.75 quartiles in transparent green fill, ensemble minimum and maximum with a dashed red line and the ensemble mean with a solid red line.}

\begin{figure}
\centering{}\subfloat[40-5]{\protect\centering{}\protect\includegraphics[width=0.3\textwidth]{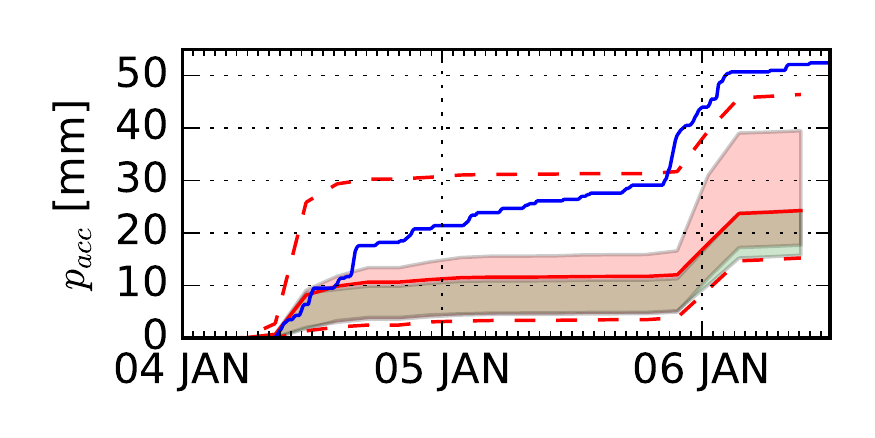}\protect}\subfloat[40-10]{\protect\centering{}\protect\includegraphics[width=0.3\textwidth]{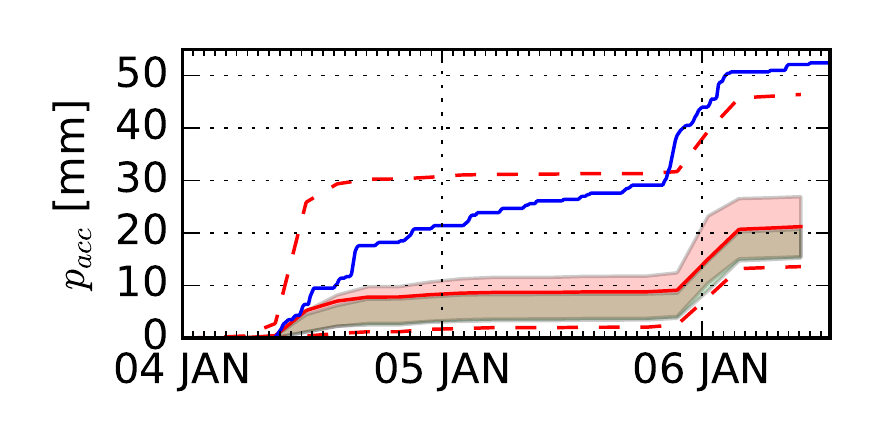}\protect}\subfloat[40-15]{\protect\centering{}\protect\includegraphics[width=0.3\textwidth]{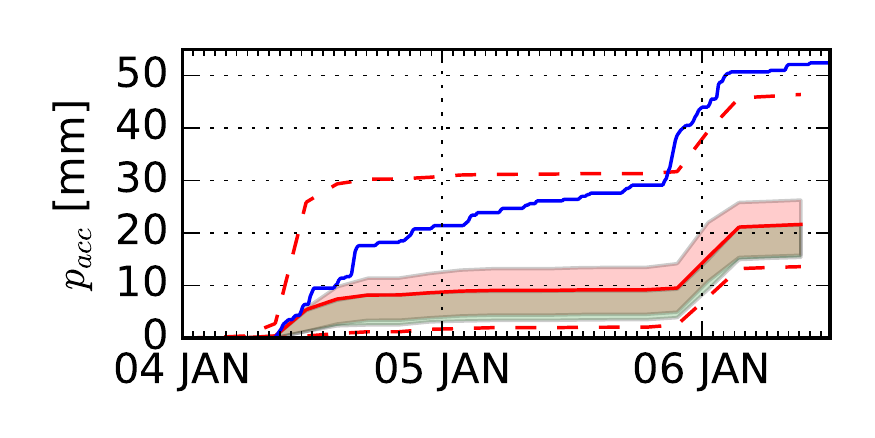}\protect}\subfloat[40-19]{\protect\centering{}\protect\includegraphics[width=0.3\textwidth]{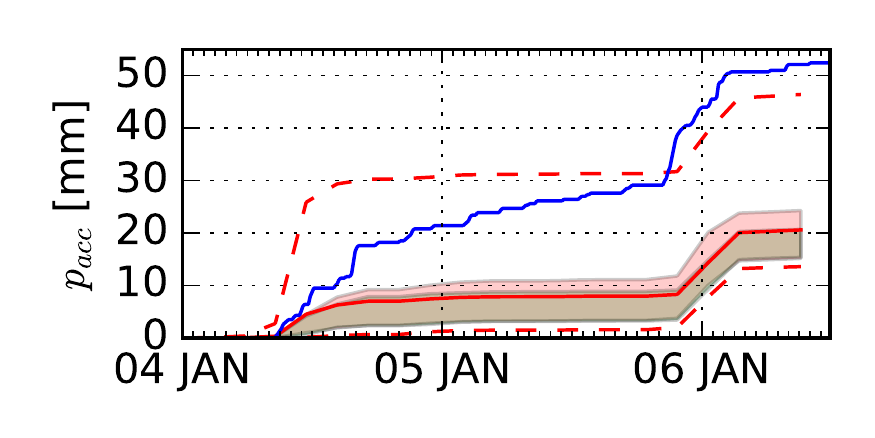}\protect}\\
\subfloat[60-5]{\protect\centering{}\protect\includegraphics[width=0.3\textwidth]{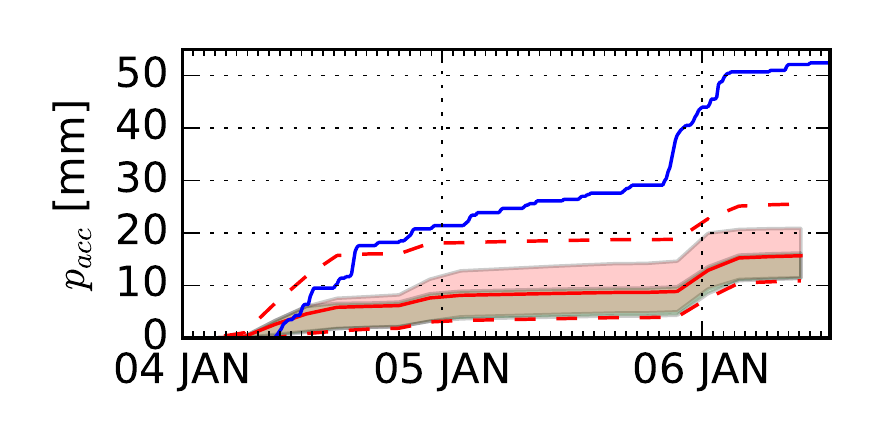}\protect}\subfloat[60-10]{\protect\centering{}\protect\includegraphics[width=0.3\textwidth]{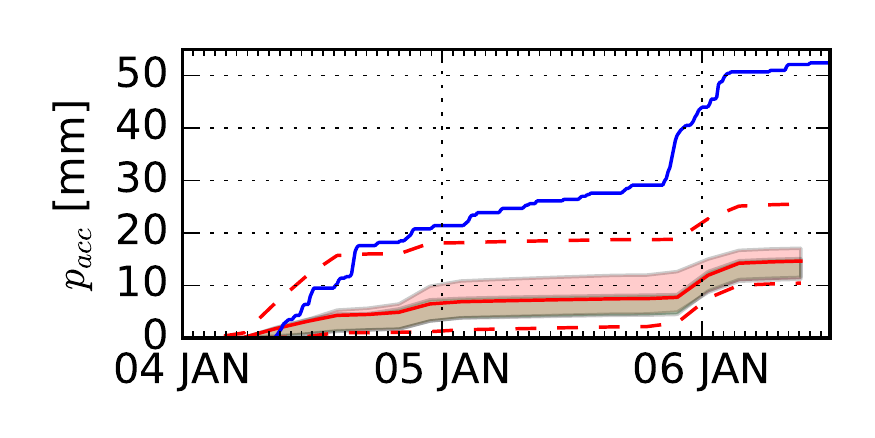}\protect}\subfloat[60-15]{\protect\centering{}\protect\includegraphics[width=0.3\textwidth]{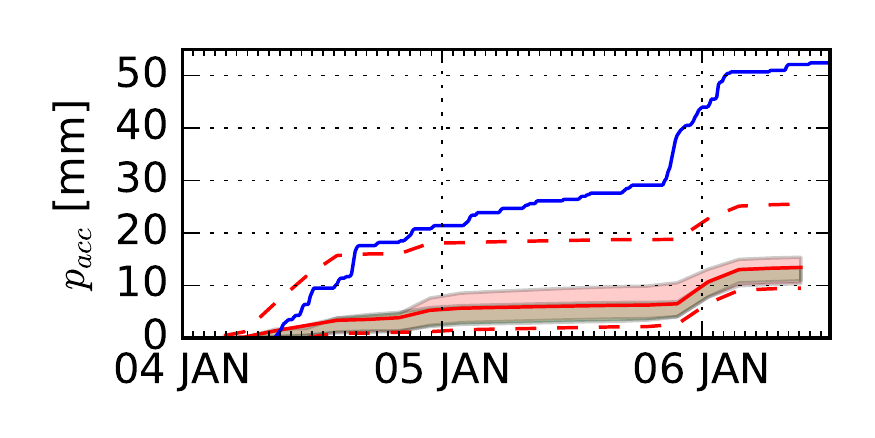}\protect}\subfloat[60-19]{\protect\centering{}\protect\includegraphics[width=0.3\textwidth]{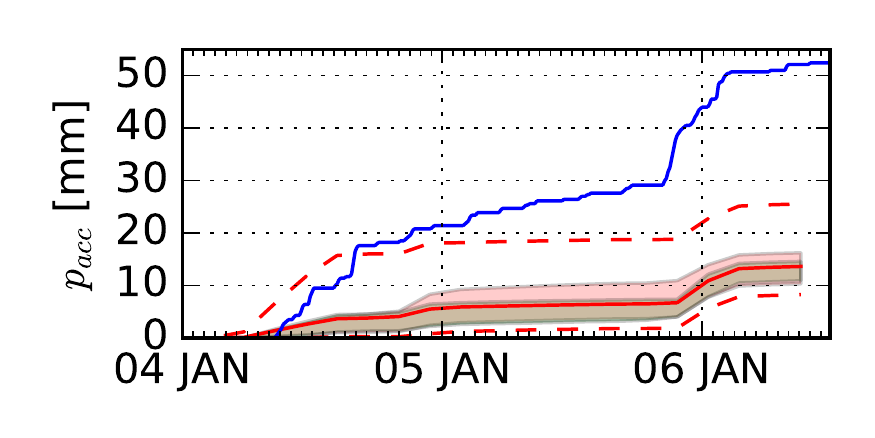}\protect}\\
\subfloat[80-5]{\protect\centering{}\protect\includegraphics[width=0.3\textwidth]{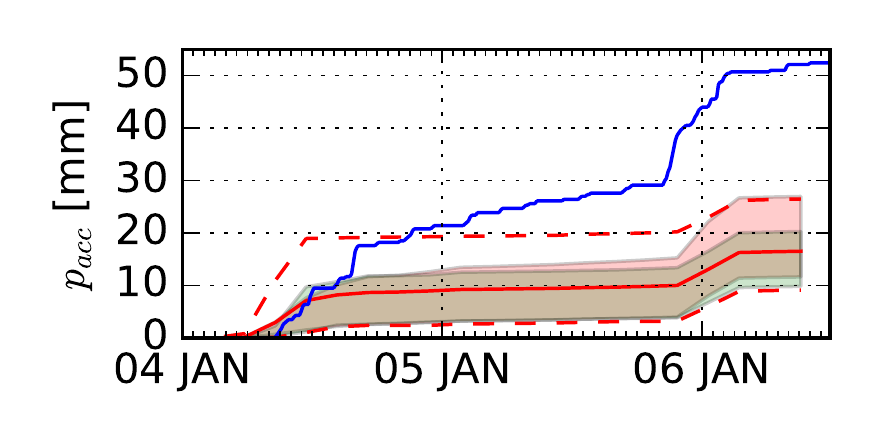}\protect}\subfloat[80-10]{\protect\centering{}\protect\includegraphics[width=0.3\textwidth]{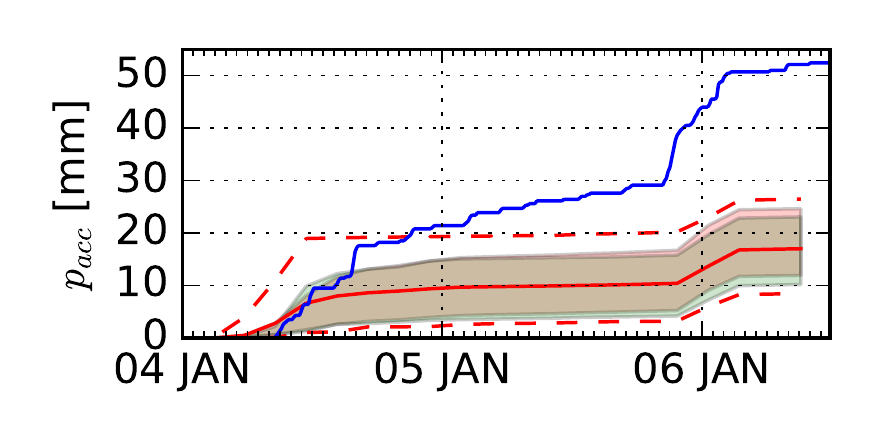}\protect}\subfloat[80-15]{\protect\centering{}\protect\includegraphics[width=0.3\textwidth]{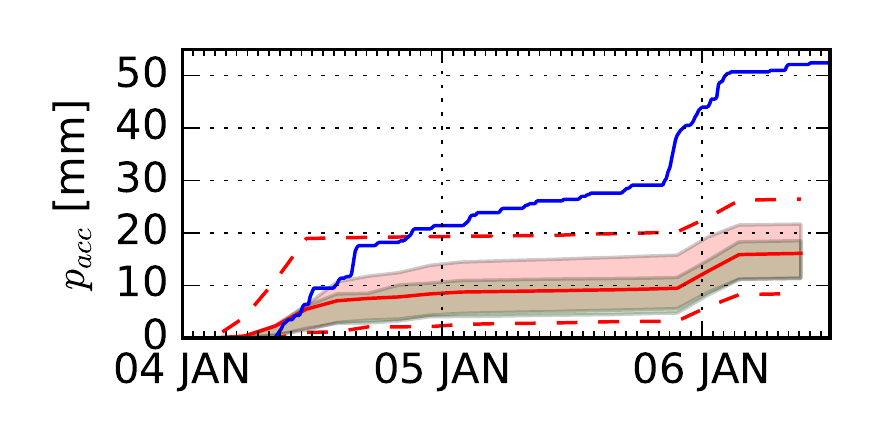}\protect}\subfloat[80-19]{\protect\centering{}\protect\includegraphics[width=0.3\textwidth]{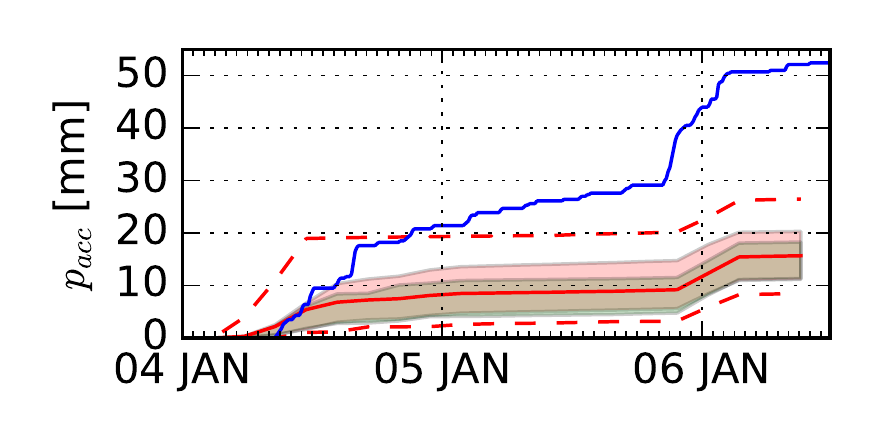}\protect}\\
\subfloat[100-5]{\protect\centering{}\protect\includegraphics[width=0.3\textwidth]{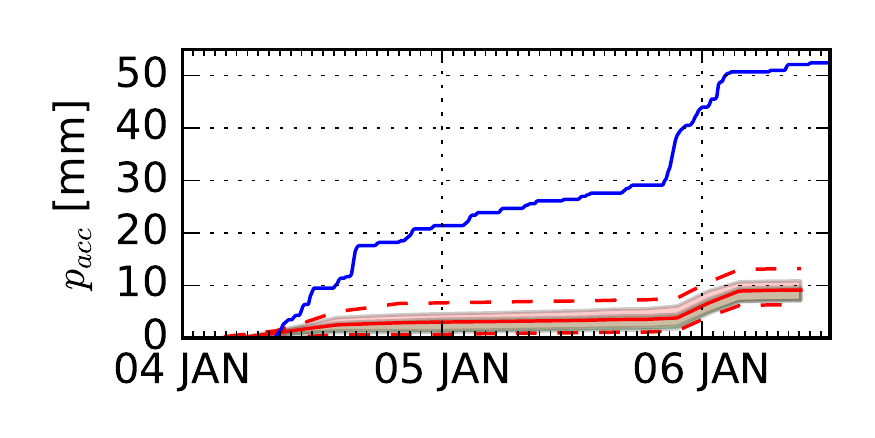}\protect}\subfloat[100-10]{\protect\centering{}\protect\includegraphics[width=0.3\textwidth]{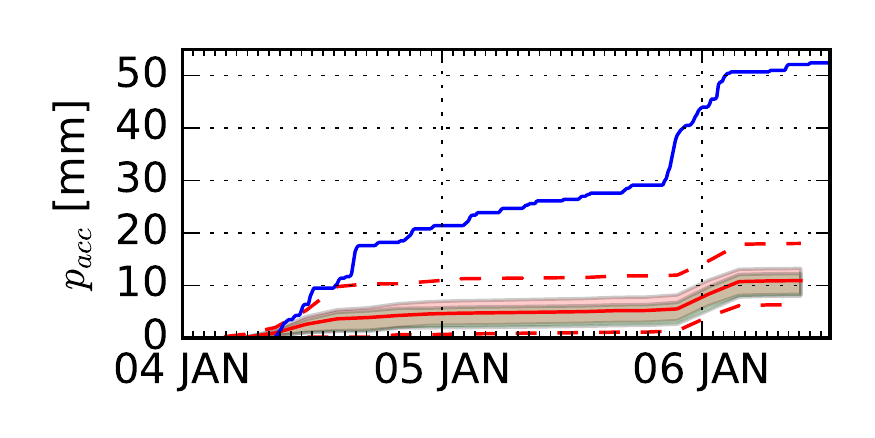}\protect}\subfloat[100-15]{\protect\centering{}\protect\includegraphics[width=0.3\textwidth]{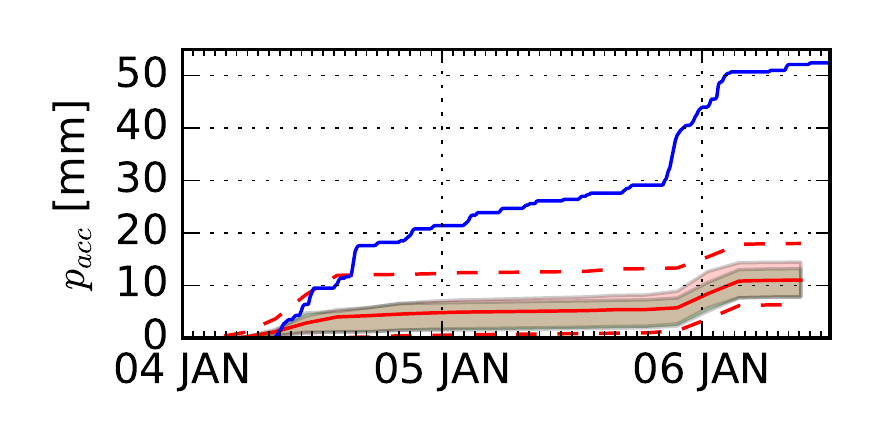}\protect}\subfloat[100-19]{\protect\centering{}\protect\includegraphics[width=0.3\textwidth]{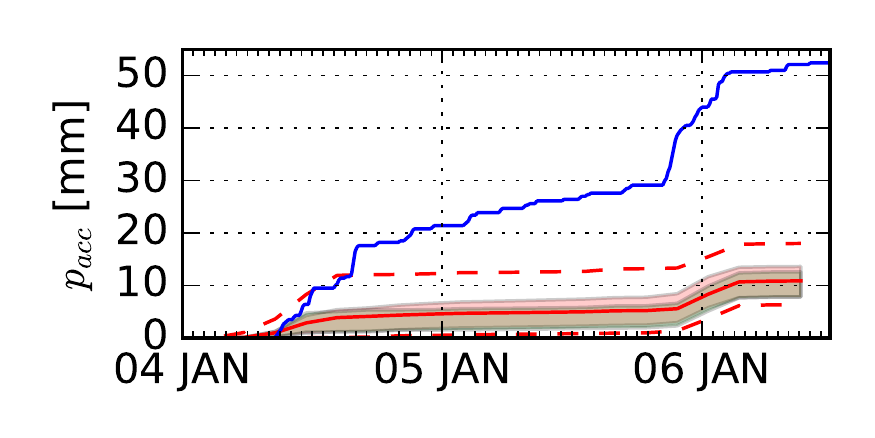}\protect}\\
\subfloat[120-5]{\protect\centering{}\protect\includegraphics[width=0.3\textwidth]{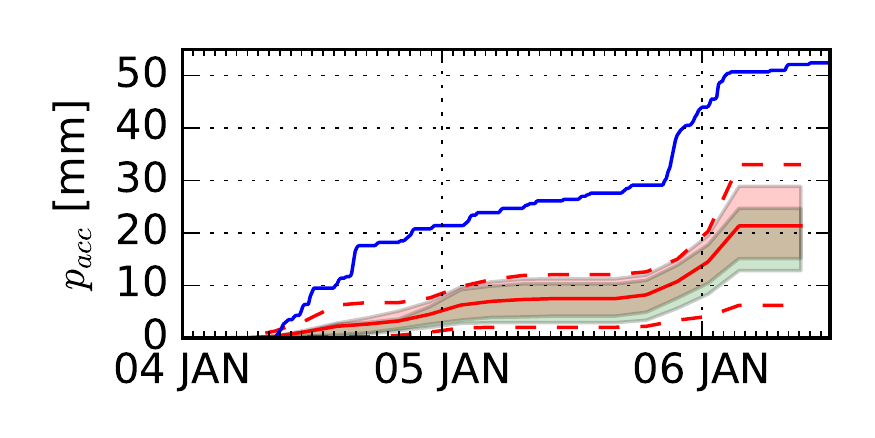}\protect}\subfloat[120-10]{\protect\centering{}\protect\includegraphics[width=0.3\textwidth]{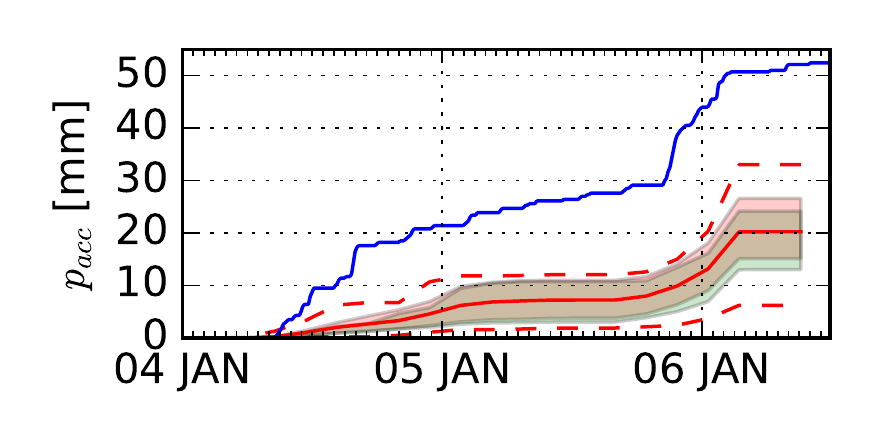}\protect}\subfloat[120-15]{\protect\centering{}\protect\includegraphics[width=0.3\textwidth]{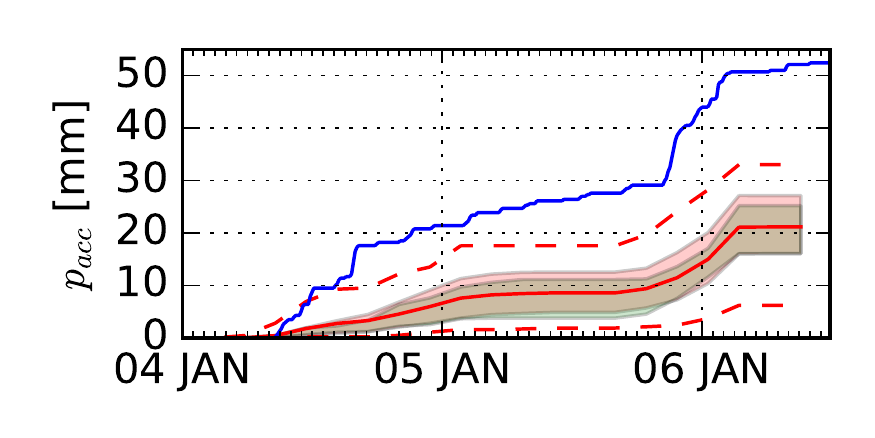}\protect}\subfloat[120-19]{\protect\centering{}\protect\includegraphics[width=0.3\textwidth]{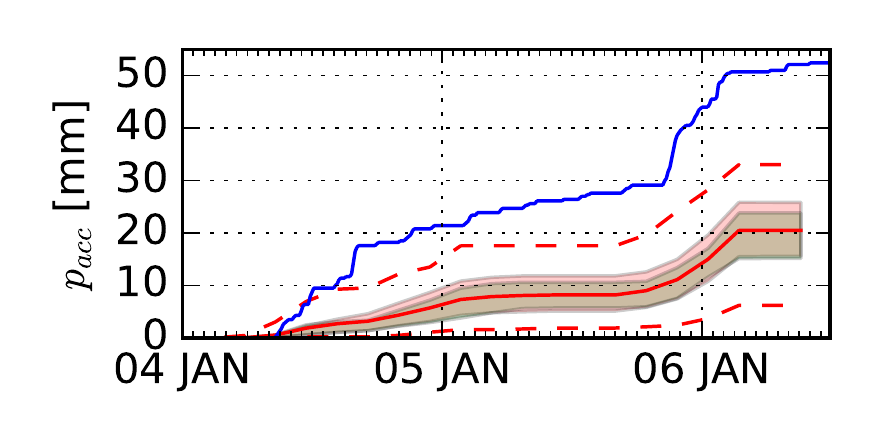}\protect}\\
\subfloat[160-5]{\protect\centering{}\protect\includegraphics[width=0.3\textwidth]{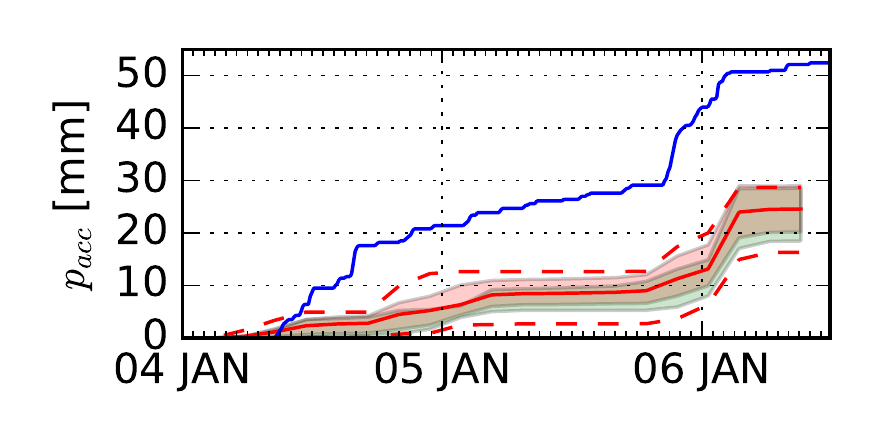}\protect}\subfloat[160-10]{\protect\centering{}\protect\includegraphics[width=0.3\textwidth]{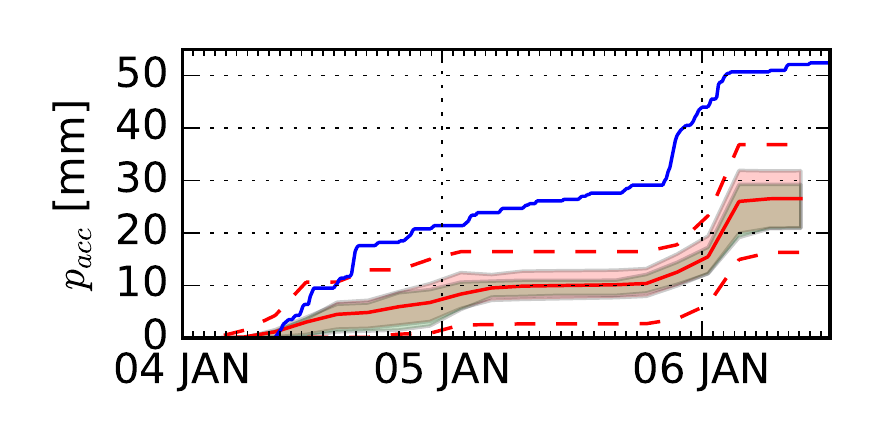}\protect}\subfloat[160-15]{\protect\centering{}\protect\includegraphics[width=0.3\textwidth]{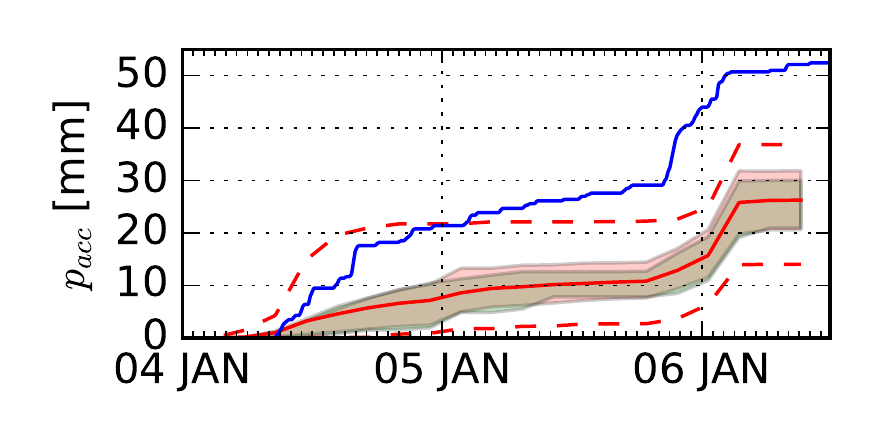}\protect}\subfloat[160-19]{\protect\centering{}\protect\includegraphics[width=0.3\textwidth]{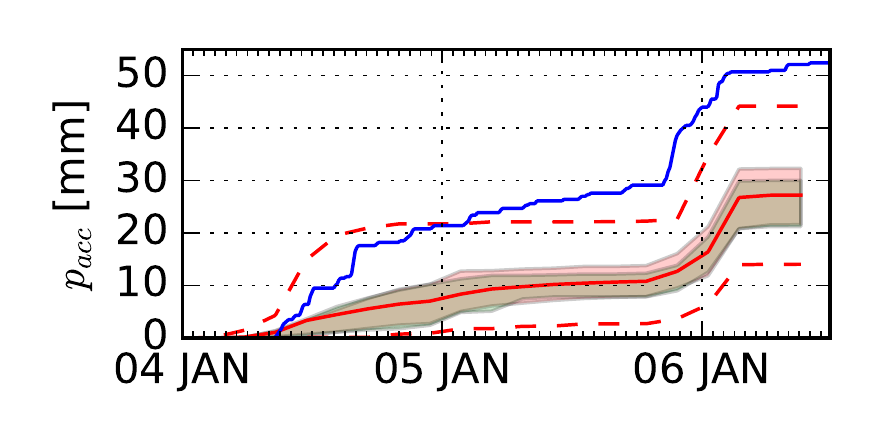}\protect}\\
\protect\caption{Accumulated precipitation, $p_{acc}$, for initial time: 04 Jan 2016 00:00 UTC, domain d03.
Dashed lines: Minimum and maximum across all ensemble members,  cumulative density function interval from 0.25 until 0.75 in red
shading and quartiles in green shading. Observations are plotted in blue and ensemble mean in red.
The caption of the sub panels indicated domain size followed by ensemble size (members considered). \label{fig:Jan-04}}
\end{figure}

\begin{figure}
\centering{}\subfloat[40-5]{\protect\centering{}\protect\includegraphics[width=0.3\textwidth]{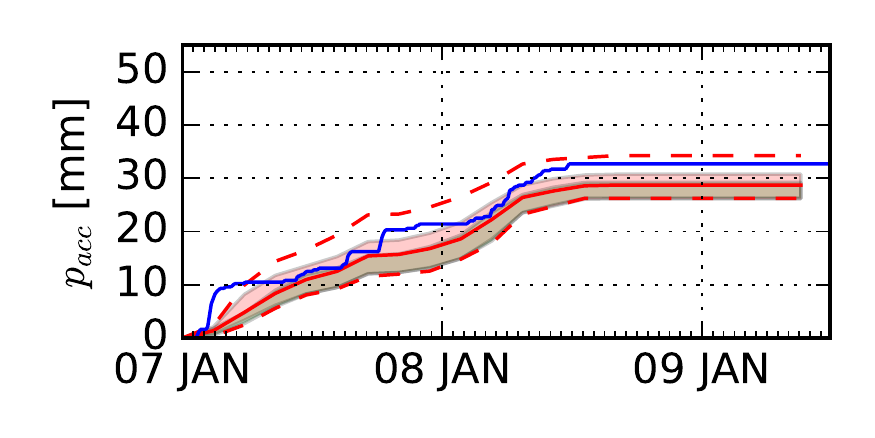}\protect}\subfloat[40-10]{\protect\centering{}\protect\includegraphics[width=0.3\textwidth]{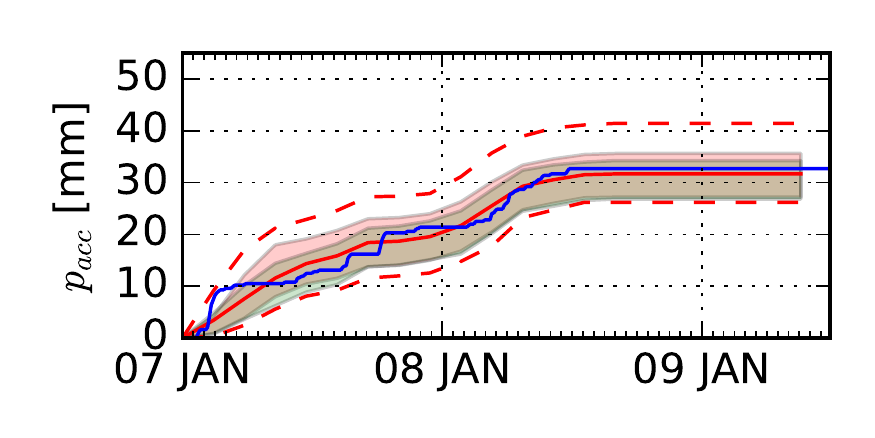}\protect}\subfloat[40-15]{\protect\centering{}\protect\includegraphics[width=0.3\textwidth]{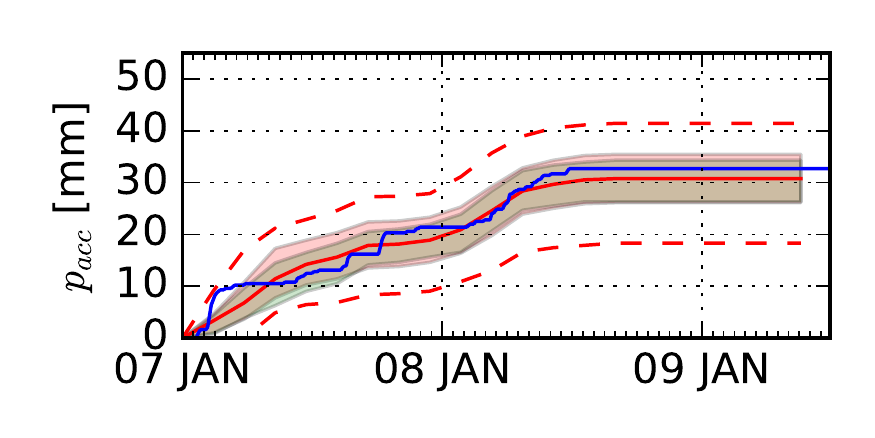}\protect}\subfloat[40-19]{\protect\centering{}\protect\includegraphics[width=0.3\textwidth]{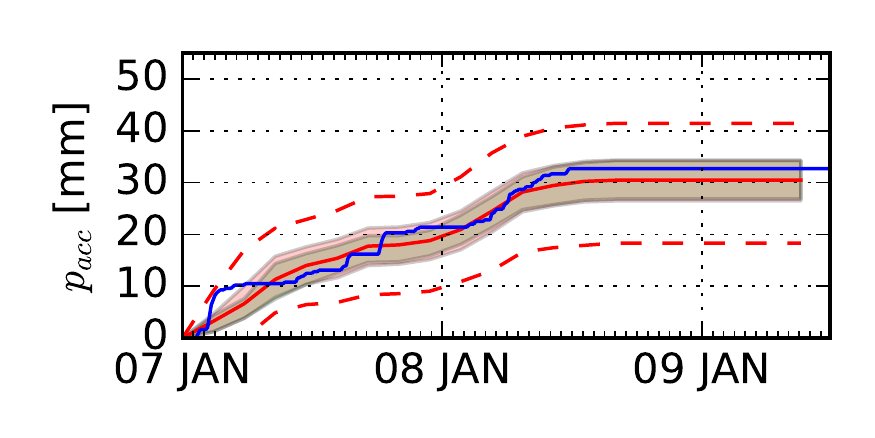}\protect}\\
\subfloat[60-5]{\protect\centering{}\protect\includegraphics[width=0.3\textwidth]{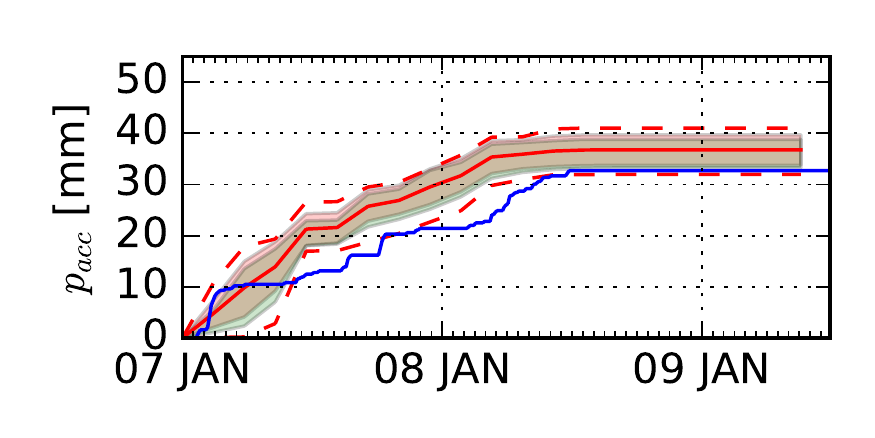}\protect}\subfloat[60-10]{\protect\centering{}\protect\includegraphics[width=0.3\textwidth]{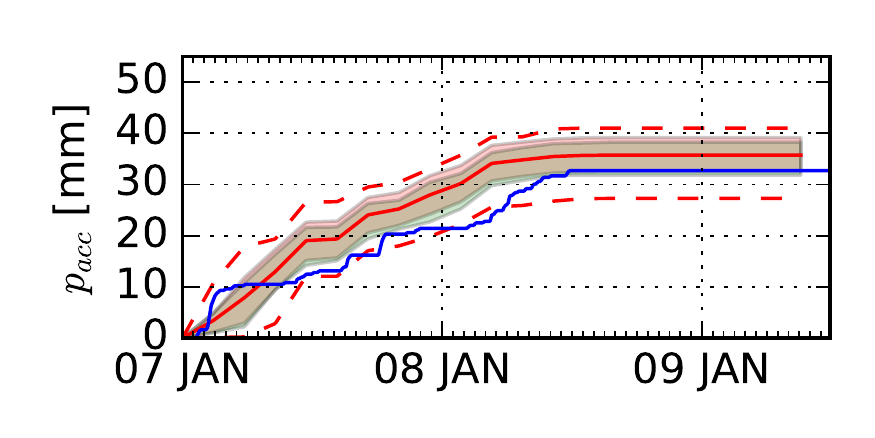}\protect}\subfloat[60-15]{\protect\centering{}\protect\includegraphics[width=0.3\textwidth]{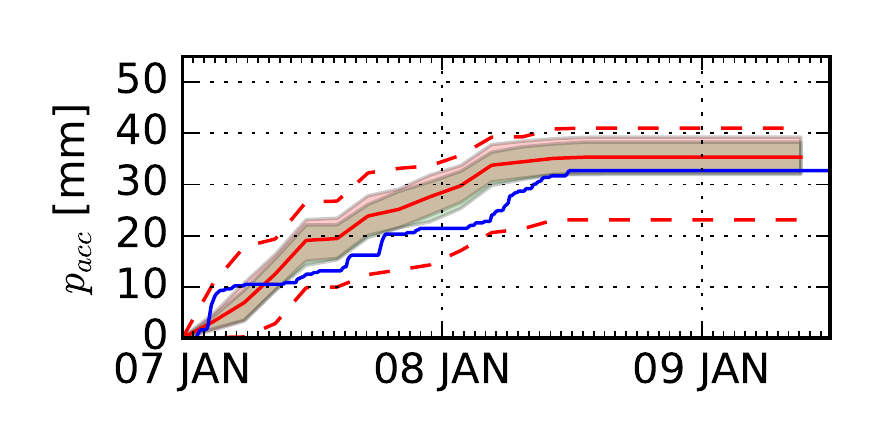}\protect}\subfloat[60-19]{\protect\centering{}\protect\includegraphics[width=0.3\textwidth]{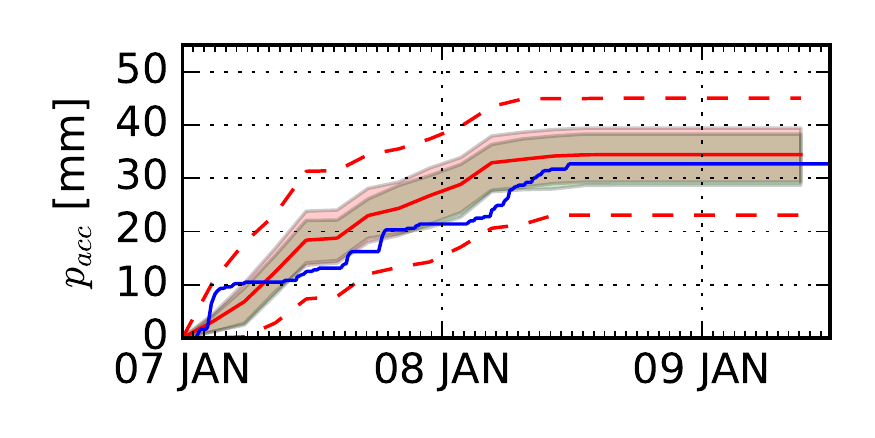}\protect}\\
\subfloat[80-5]{\protect\centering{}\protect\includegraphics[width=0.3\textwidth]{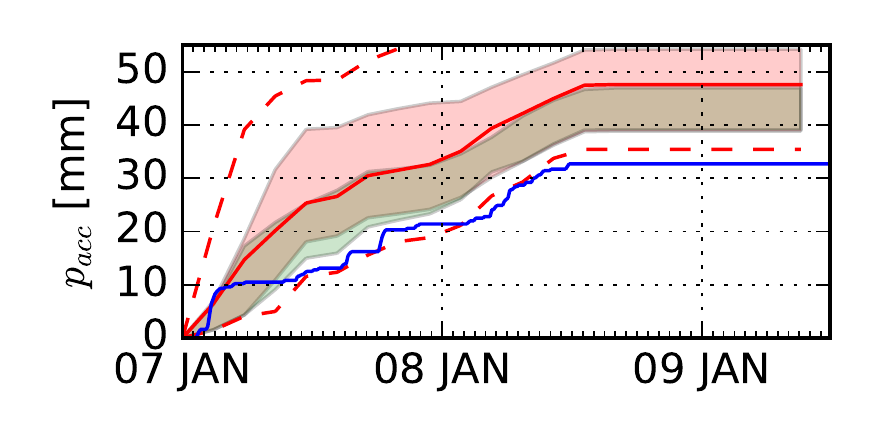}\protect}\subfloat[80-10]{\protect\centering{}\protect\includegraphics[width=0.3\textwidth]{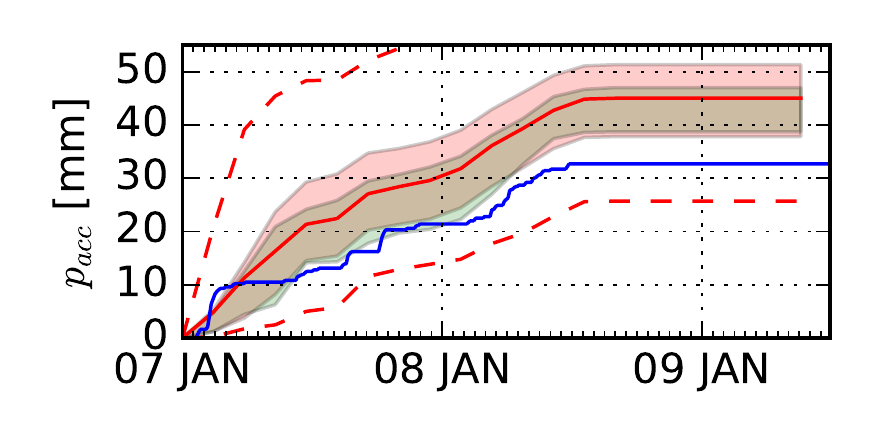}\protect}\subfloat[80-15]{\protect\centering{}\protect\includegraphics[width=0.3\textwidth]{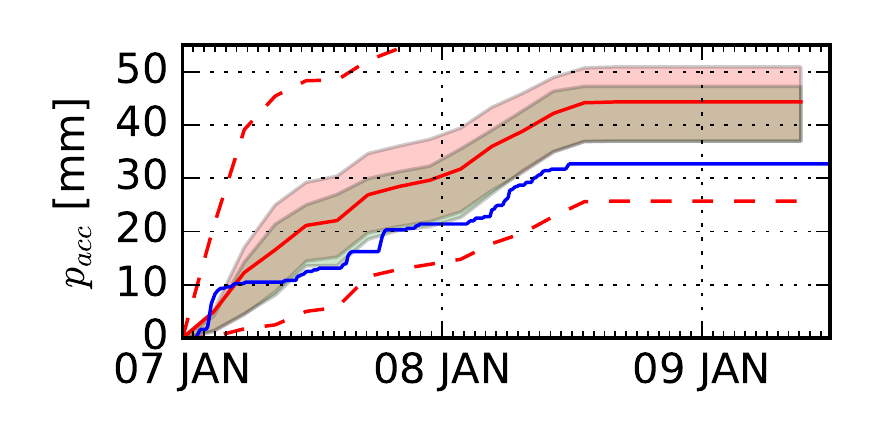}\protect}\subfloat[80-19]{\protect\centering{}\protect\includegraphics[width=0.3\textwidth]{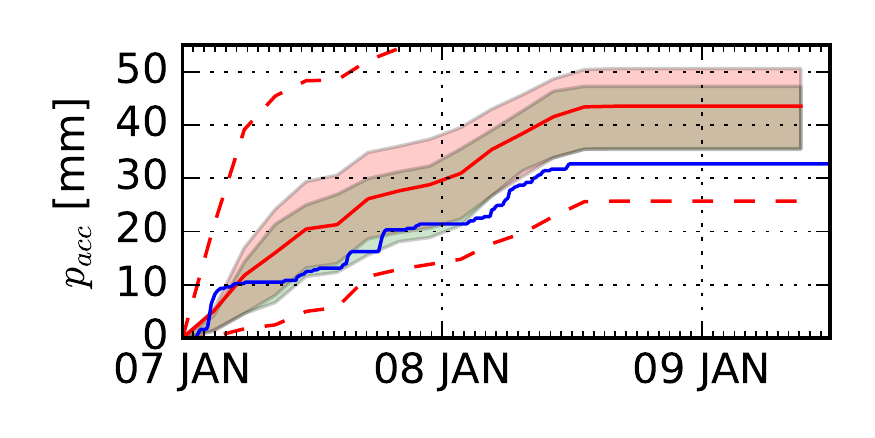}\protect}\\
\subfloat[100-5]{\protect\centering{}\protect\includegraphics[width=0.3\textwidth]{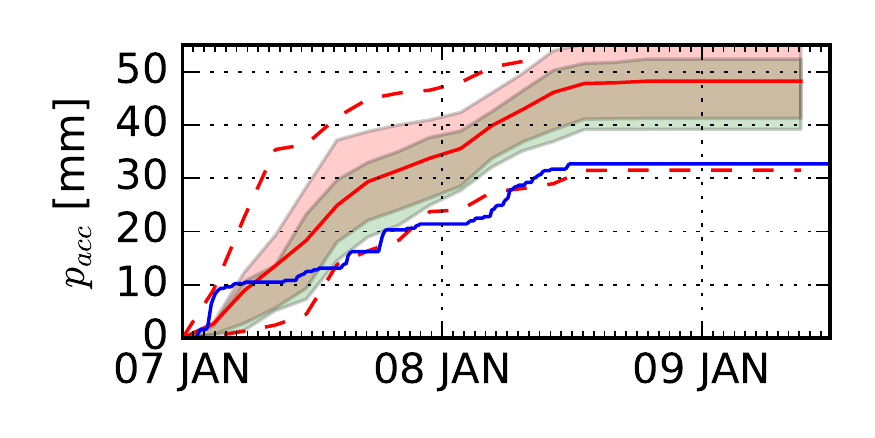}\protect}\subfloat[100-10]{\protect\centering{}\protect\includegraphics[width=0.3\textwidth]{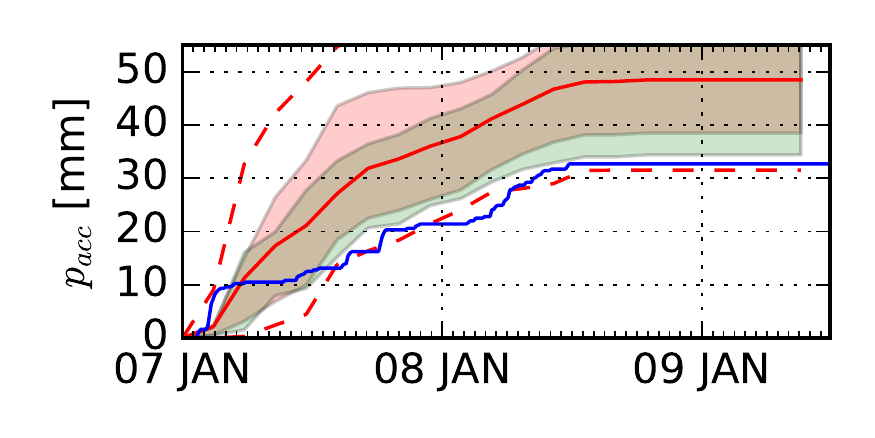}\protect}\subfloat[100-15]{\protect\centering{}\protect\includegraphics[width=0.3\textwidth]{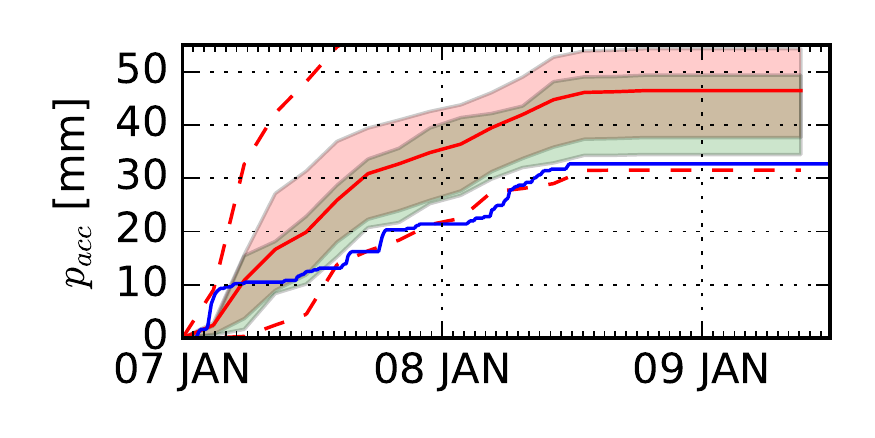}\protect}\subfloat[100-19]{\protect\centering{}\protect\includegraphics[width=0.3\textwidth]{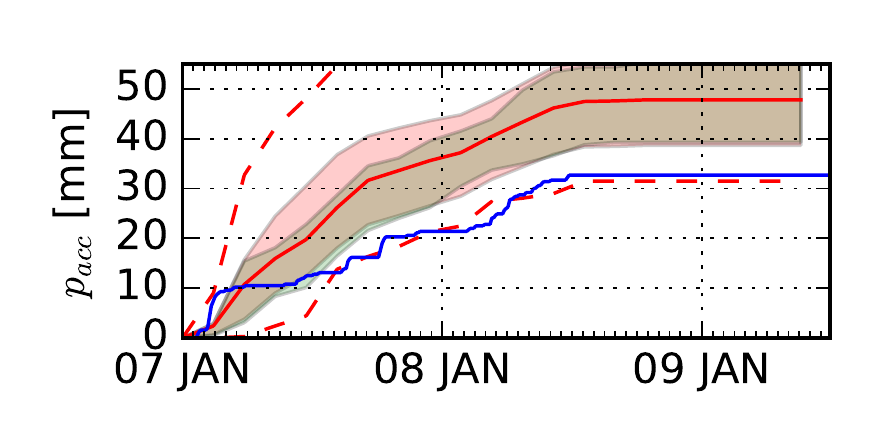}\protect}\\
\subfloat[120-5]{\protect\centering{}\protect\includegraphics[width=0.3\textwidth]{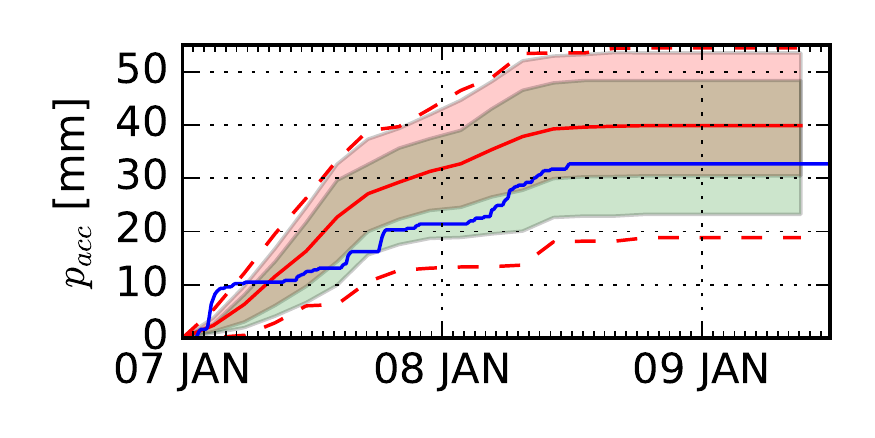}\protect}\subfloat[120-10]{\protect\centering{}\protect\includegraphics[width=0.3\textwidth]{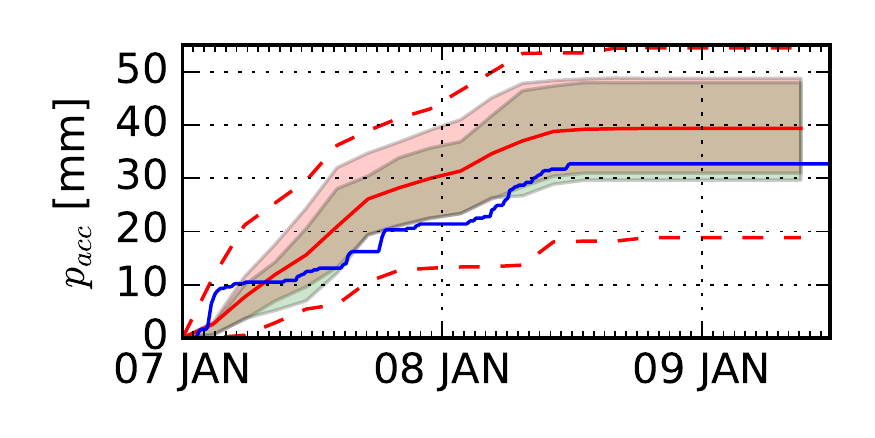}\protect}\subfloat[120-15]{\protect\centering{}\protect\includegraphics[width=0.3\textwidth]{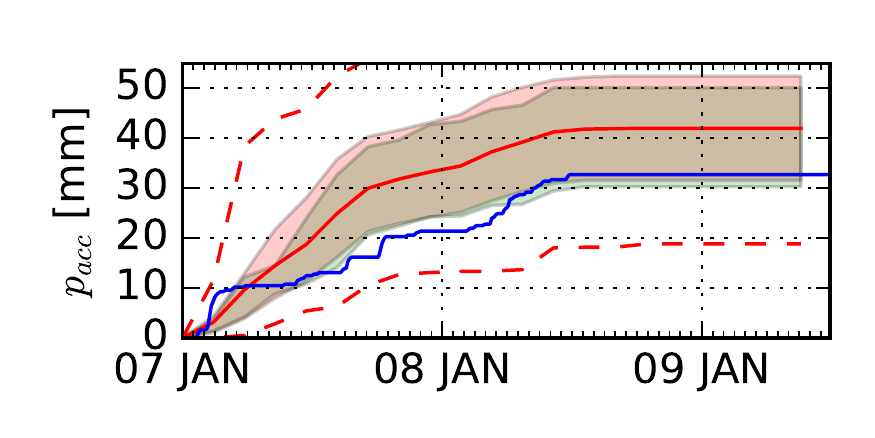}\protect}\subfloat[120-19]{\protect\centering{}\protect\includegraphics[width=0.3\textwidth]{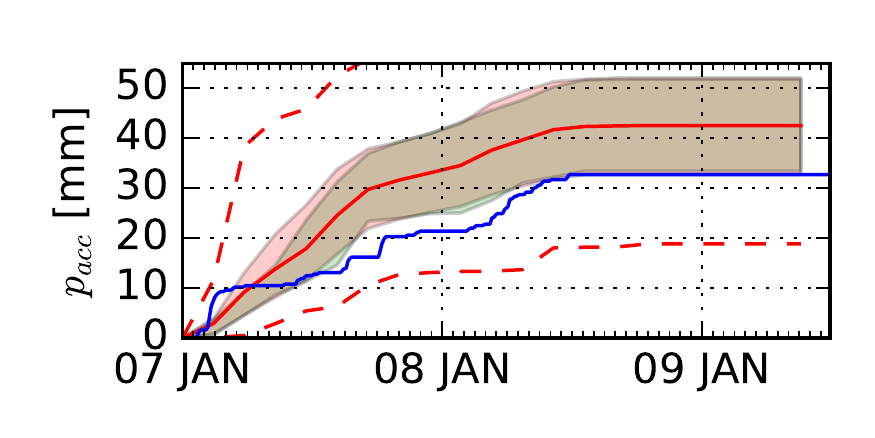}\protect}\\
\subfloat[160-5]{\protect\centering{}\protect\includegraphics[width=0.3\textwidth]{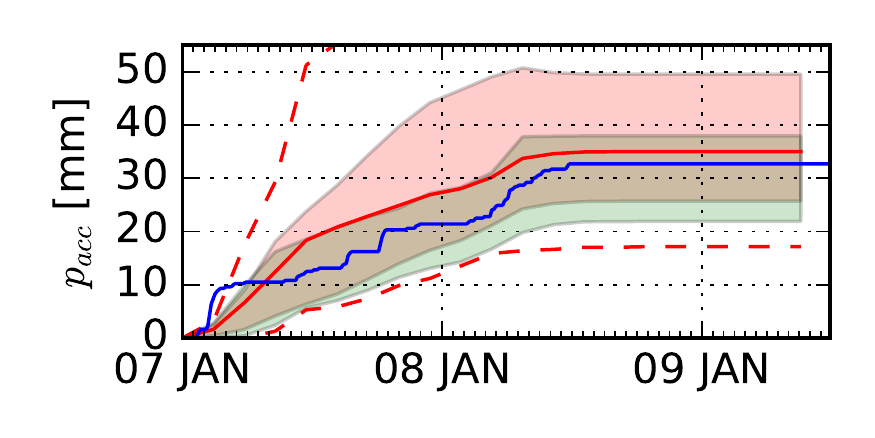}\protect}\subfloat[160-10]{\protect\centering{}\protect\includegraphics[width=0.3\textwidth]{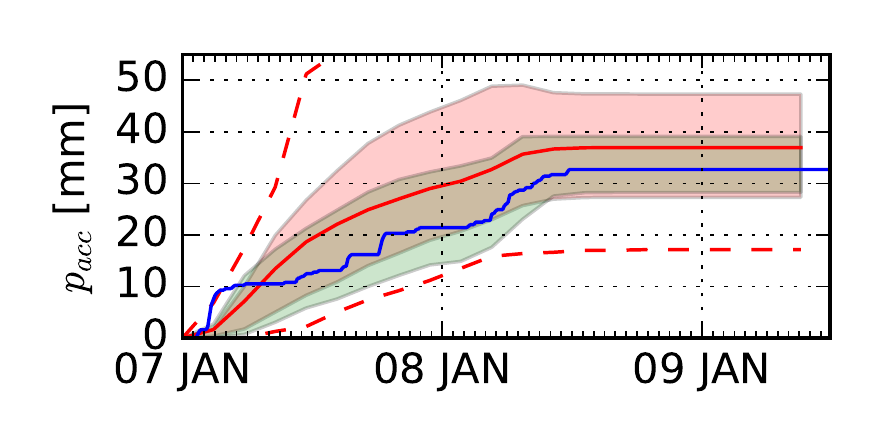}\protect}\subfloat[160-15]{\protect\centering{}\protect\includegraphics[width=0.3\textwidth]{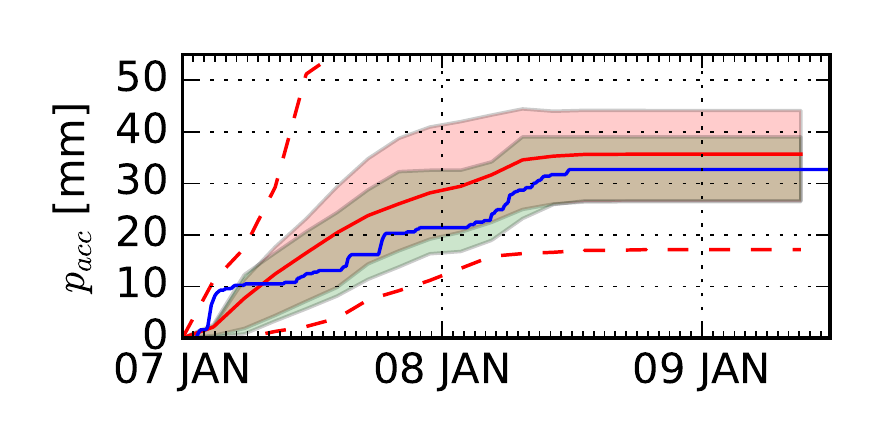}\protect}\subfloat[160-19]{\protect\centering{}\protect\includegraphics[width=0.3\textwidth]{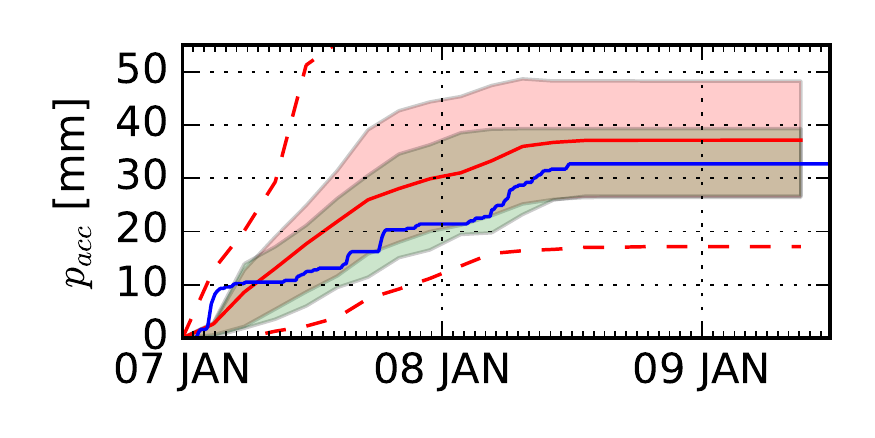}\protect}\\
\protect\caption{Accumulated precipitation, $p_{acc}$, for initial time: 07 Jan 2016 00:00 UTC, domain d03.
Dashed lines: Minimum and maximum across all ensemble members,  cumulative density function interval from 0.25 until 0.75  in red
shading and quartiles in green shading. Observations are plotted in blue and ensemble mean in red.
The caption of the sub panels indicated domain size followed by ensemble size (members considered). \label{fig:Jan07}}
\end{figure}

In Figures \ref{fig:Jan-04} and \ref{fig:Jan07} the time series
of accumulated precipitation are plotted. Each of the plots shows
six components: Minimum and maximum across all ensemble members, ensemble
mean (red), 25th and 75th quartile for the cpd (c.f. Equation \ref{eq:cpd})
shown in red shading, 25th and 75th quartile for the interpolation
approach (c.f. Equation \ref{eq:trad}) shown in green shading and
the observations (blue). The sub panels present the information for
the different domain sizes, n=40 to 160, top to bottom, and
number of ensemble members considered, five to 19, left to right.
\subsubsection{Initial time 04 January 2016 00:00 UTC}
For initial time 04 January 2016 00:00 UTC the forecasts appear to not capture the high amount of rainfall
and it is difficult to derive a clear domain size and ensemble member
count sensitivity. Interestingly, the medium domain sizes are outperformed by both, the smaller and larger domains. It may appear here as if all the ensemble members for 04 January were missing the real
weather, certainly in the ensemble mean. It will be shown later that this is an effect of the meaning. Here
the ensemble certainly was improved by adding more members and increasing
the domain size. It is not clear if increasing the domain much more
(ignoring for a while that the NAM domain does not extend much more
into the region) and adding significantly more members will eventually
improve on the second half of the forecast. Whilst there is improvement
when comparing the panels in Figure~\ref{fig:Jan-04}, the improvement  is small.

\subsubsection{Initial time 07 January 2016 00:00 UTC}
The results are different for the 07 January 2016 00:00 UTC data set. There
it can be readily observed how the ensemble is overly constrained for
the small domain sizes by its lateral boundary conditions (cf. \citet{QJ:QJ2238}),
which are unperturbed. For the small domain sizes the ensemble spread, the difference between
minimum and maximum, is small. Coincidentally,
for the initial time of 07 January 2016 00:00 UTC the results are close to the observations,
however, this in itself does not make them more likely. With the low
spread of the small domain ensemble it simply means that the mesoscale
forecast was already close to observations. The spread increases as the domain
size passes $n=80$. With the ensemble mean staying close
to the observations this then increases the confidence.

\subsection{Spacial uncertainty\label{sub:Spacial-uncertainty}}

\begin{figure}
\begin{centering}
\subfloat[Accumulated precipitation at the end of the $62$hr forecast for the largest domain ($n=160$) with highest resolution (d03).]{\protect\centering{}\protect\includegraphics[width=0.6\textwidth]{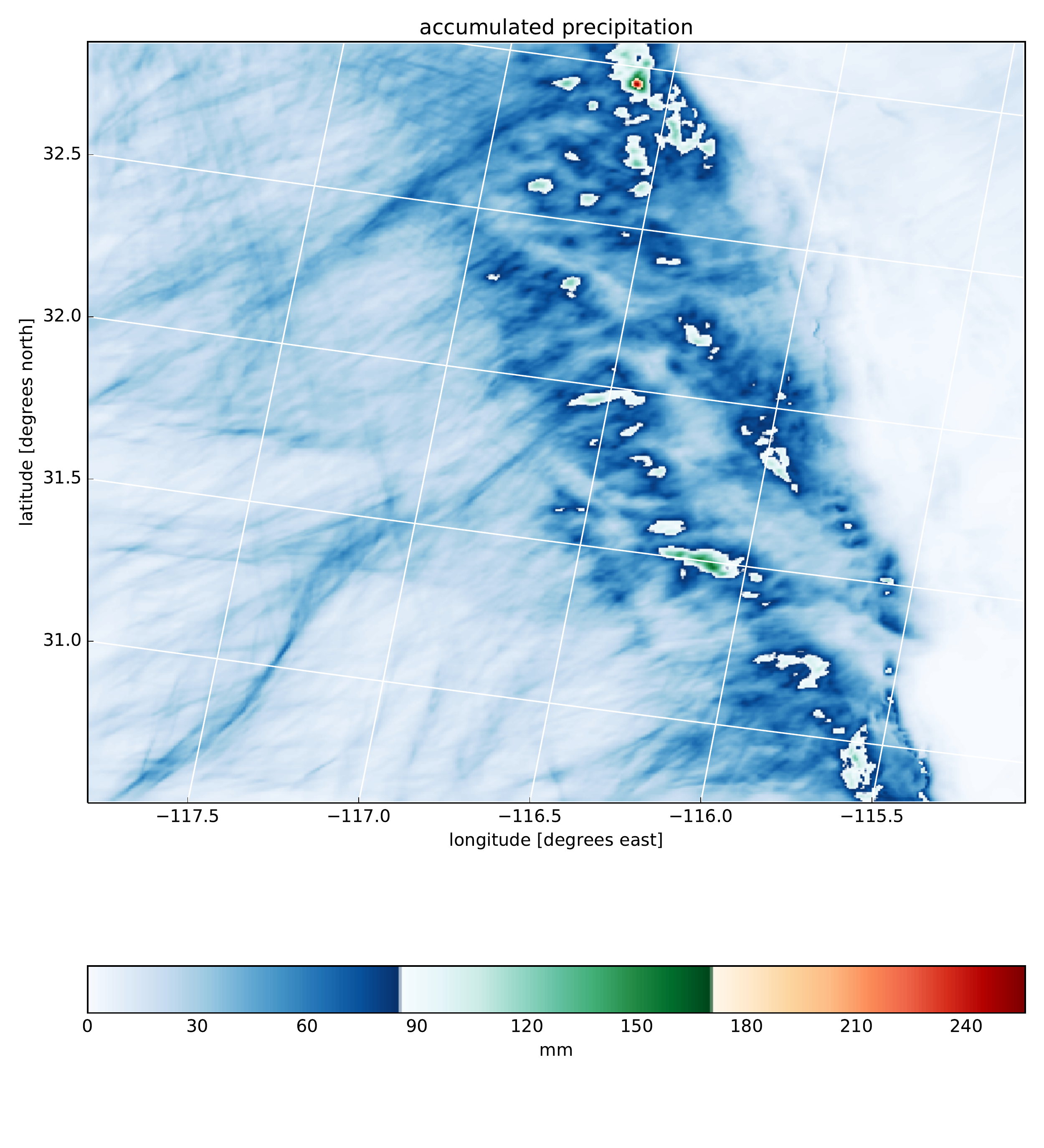}\protect}\\
\subfloat[Terrain in the highest resolution domain]{\protect\centering{}\protect\includegraphics[width=0.6\textwidth]{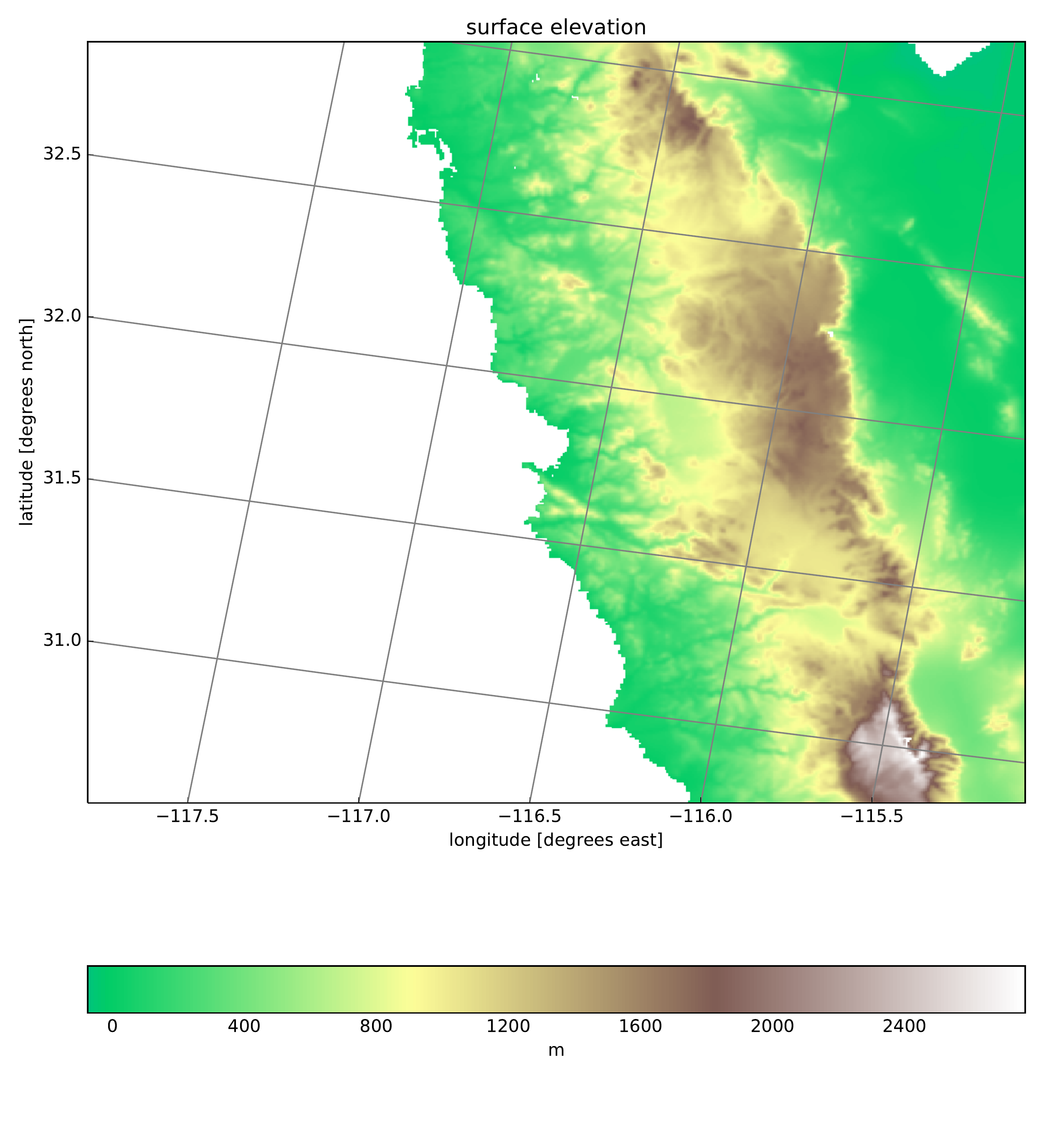}\protect}\protect\caption{Spatial distribution of precipitation\label{fig:Spatial-distribution-of-precip}}
\par\end{centering}
\end{figure}

The two examples above showed that the ensemble alone is not always sufficient. To improve
the forecast there are two options: Increase the ensemble members
or utilize the spatial variation. Here increasing the ensemble members does not appear as promising, in particular since the aim is to reduce the computational effort. 

 Since the computational effort of the downscaling increases linearly but the spatial sampling available increases quadratically the \textit{subgrid ensemble enhanced mesoscale forecast} is more cost effective.

  Figure~\ref{fig:Spatial-distribution-of-precip} shows that the spatial distribution of the precipitation is indeed highly non-uniform. A spatial sampling error therefore can have a large impact on the forecast accuracy.
\begin{figure}
\begin{centering}
\subfloat[Subgrid enhanced]{\protect\centering{}\protect\includegraphics[width=0.8\textwidth]{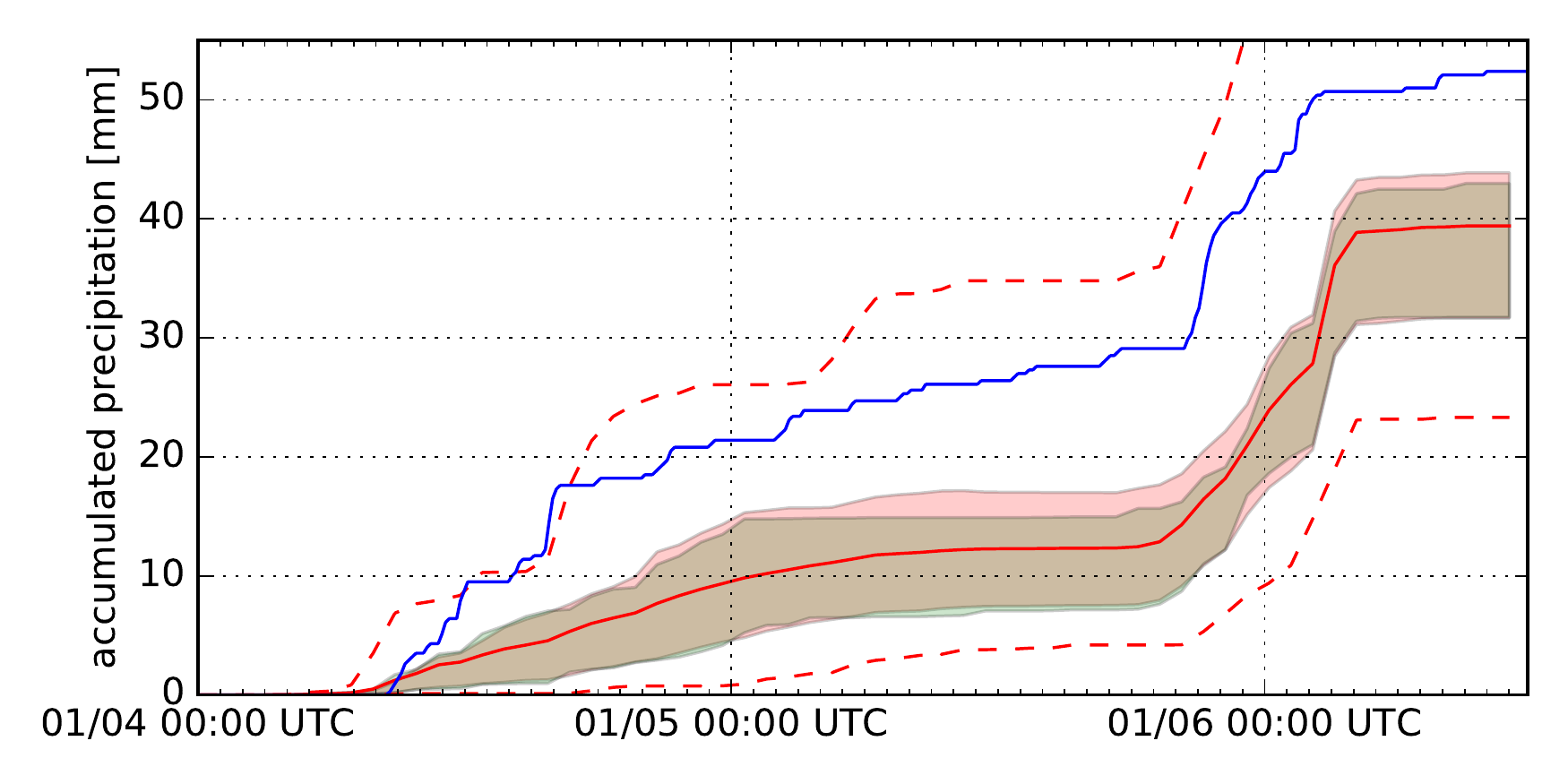}\protect}\\
\subfloat[Point ensemble]{\protect\centering{}\protect\includegraphics[width=0.8\textwidth]{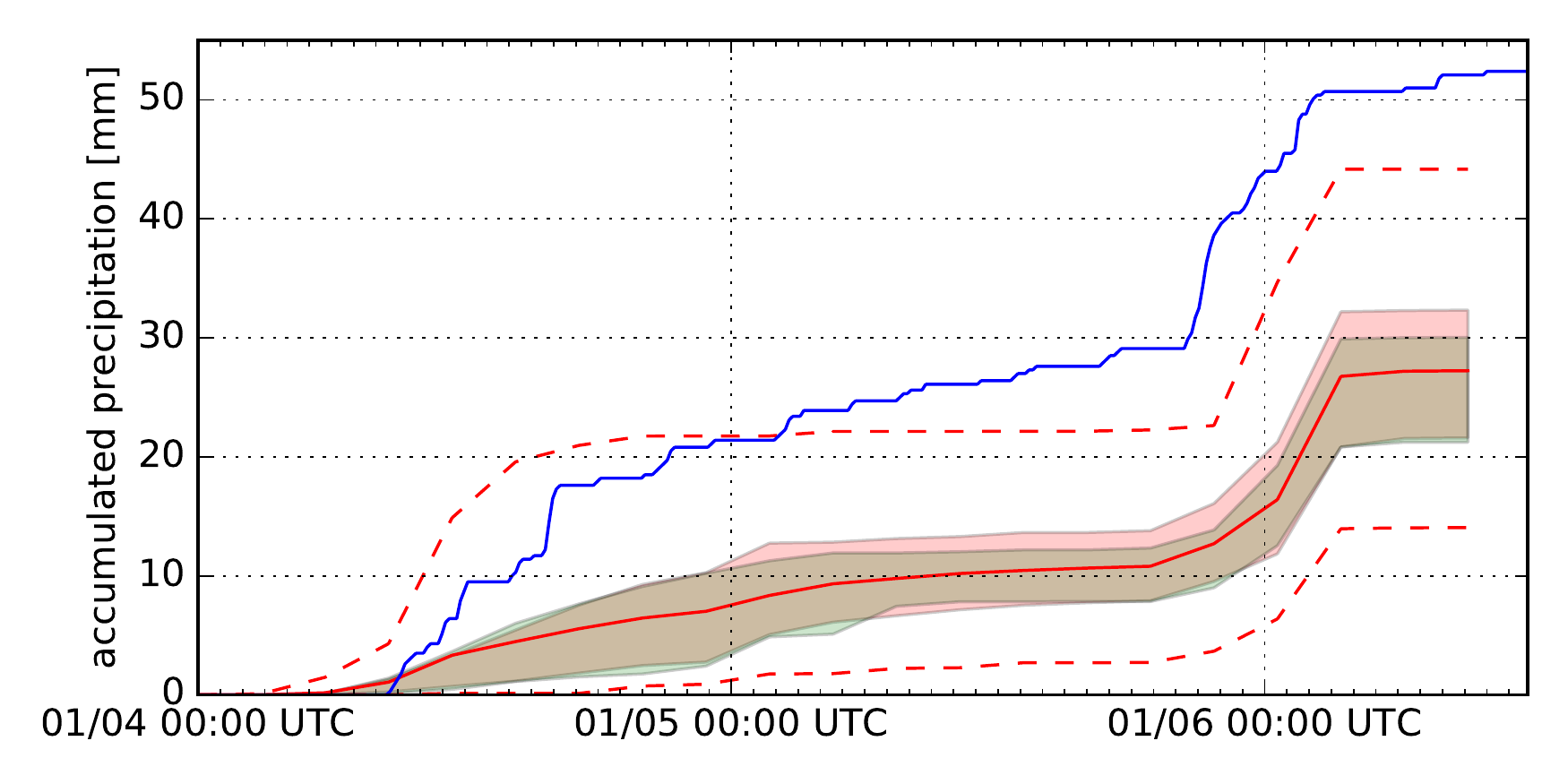}\protect}\protect\caption{Precipitation quartiles for the large domain ensemble. initial time: 04 Jan 2016 00:00 UTC. The observed precipitation is plotted in blue, the cumulative density function interval from 0.25 until 0.75 in transparent red fill, 0.25 to 0.75 quartiles in transparent green fill, ensemble minimum and maximum with a dashed red line and the ensemble mean with a solid red line.\label{fig:Precipitation-quartiles-for-subgrid} }

\par\end{centering}

\end{figure}
Figure \ref{fig:Precipitation-quartiles-for-subgrid} shows the effect
of  $3\times 3$ subgrid sampling. The sampling generates a $9$ grid point ensemble with $20\times9=180$
members and compares it to the non-sample, single point $20$ member
ensemble for the forecast run with initial date 04 January 2016, the
most difficult case due to its extreme precipitation. It can be seen
that the subgrid sampling has clearly improved the forecast. The $25$\%
and $75$\% quartiles have moved up, in the direction of the observations
and the minimum and maximum now nearly always incorporate the observations.
Overall, however, the precipitation for this extreme case is still
underpredicted by a factor of about two.

\subsection{Modal analysis: Extract potential branches\label{sub:Modal-analysis:-Extract}}

\begin{figure}
\includegraphics[width=1\textwidth]{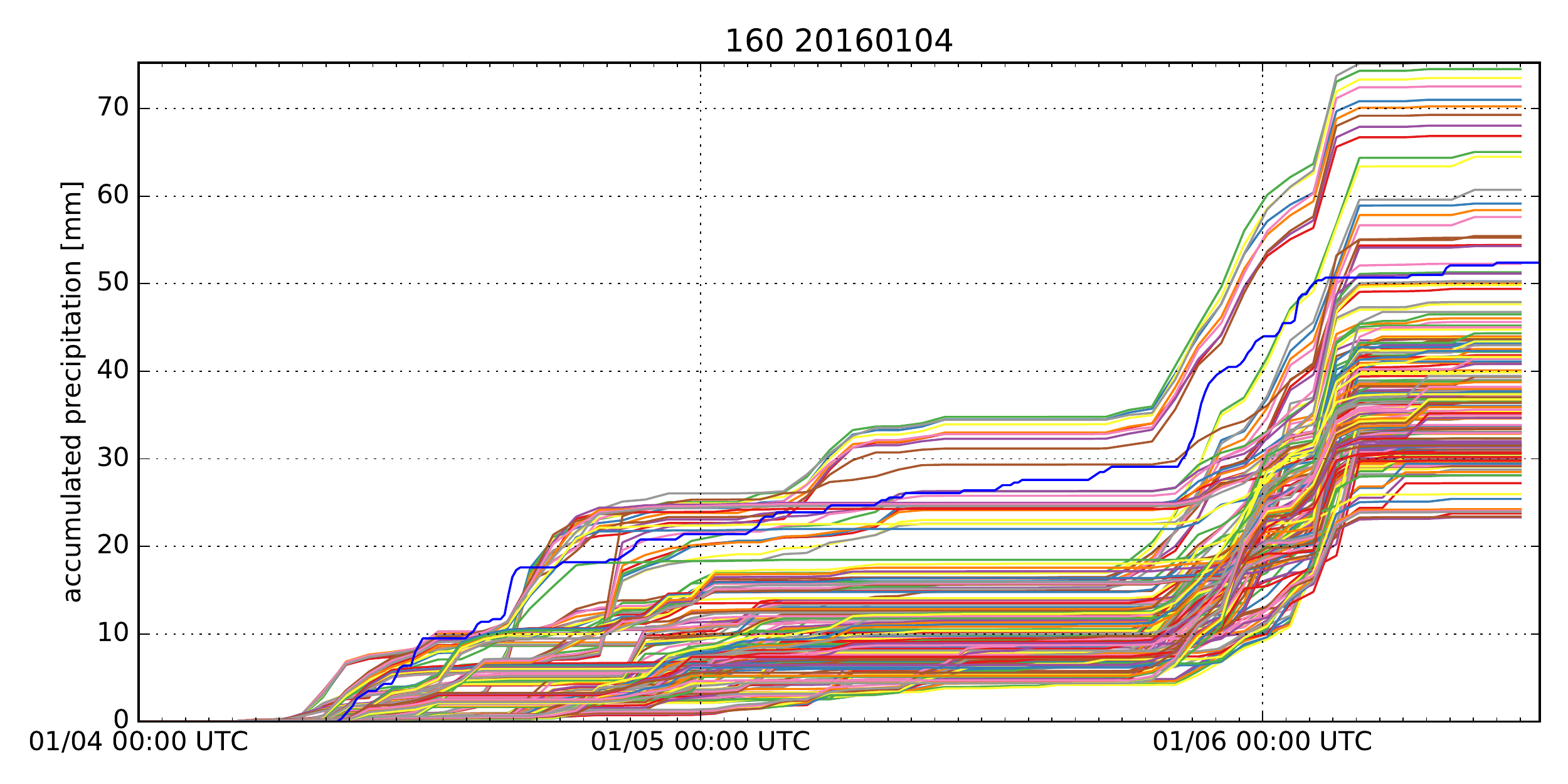}\protect\caption{Geographic and perturbation ensemble. Initial time: 04 Jan 2016 00:00 UTC. The
blue line represents the observations in grid point $i,j$ and each
color of the remaining lines represents a unique geographical location
$\left(ii,jj\right)$, with $i-1\leq ii\leq i+1$ and $j-1\leq jj\leq j+1$.
\label{fig:Jan-4-geographic-lines}}
\end{figure}

Clustering can be observed in the line plots of the time series for the $3\times 3$
grid point array in Figure \ref{fig:Jan-4-geographic-lines}. The time series for accumulated precipitation
has been color coded according to the location of the grid point. For example at the 4th of
January, 16:00UTC, clusters can be identified. It would not be surprising if these clusters would all belong to the same grid points. However, the color coding shows clearly this is not the case. This means that each cluster has contributions from several ensemble members of the other grid boxes. Analyzing the
pdf for local extrema extracts these clusters. This allows the separation into scenarios, i.e. several
possible predictions of high (or similar/significant) likelihood.

It can be seen clearly that some of these branches trace the observations closer than others.

\begin{figure}

\centering{}\includegraphics[width=1\textwidth]{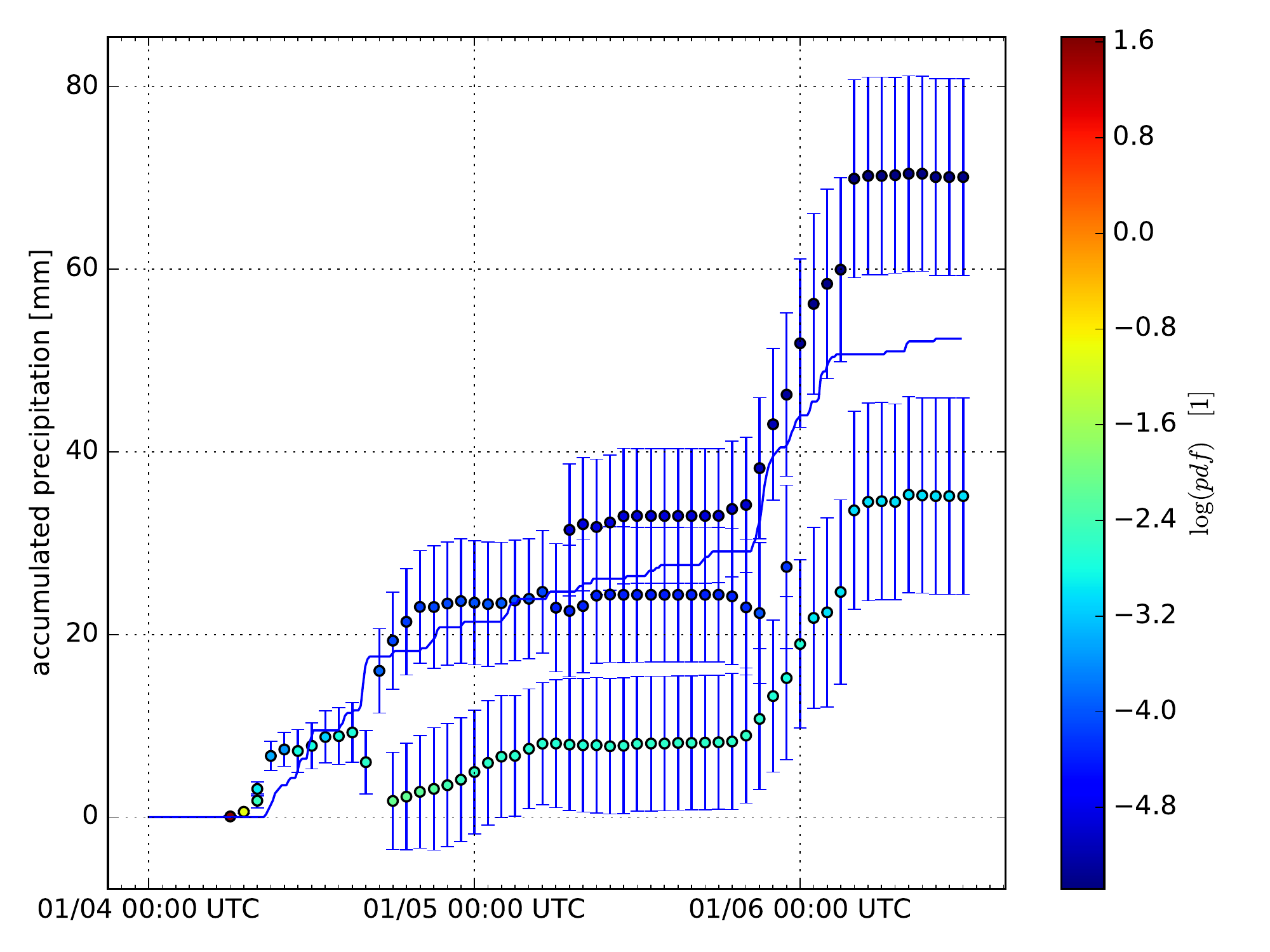}\protect\caption{Analysis of extrema of the pdf. The error bars represent one standard deviation of the ensemble members, the observed precipitation is plotted in blue. The color coding represents the logarithm
of the value of the pdf at the location of the maxima.\label{fig:Extrema-analysis}}
\end{figure}

In Figure \ref{fig:Extrema-analysis} the pdf for each time step was
analyzed for local extrema. Each data point represents a time step and extrema.
That is, if there are two points per time step there were two local maxima,
if there are three then three local maxima were observed.  Several alternative
outcomes can be observed in this \textit{extreme value analysis of
the ensemble pdf}. 
It can be seen how one of the potential outcomes comes very close
to the observed time series. For the first $24$hours the observations
are within less than one standard deviation ($\sigma$) of the forecast,
for forecast hours (fh) $24$ to $48$ within less than $\sigma/2$, focusing
on the upper branches. Initially the upper branch would not appear a sensible choice, due
to its significantly lower probability. However, as this is accumulated
precipitation it would be equally unrealistic to accept the good match
in the first $24$ hrs and then choose the lower branch for the remaining
period, which would require a reduction of accumulated precipitation
at approximately fh $18$, which is physically not possible.

\subsection{Assimilation of observations\label{sub:Assimilation-of-observations}}

In the analysis above one of the possible scenarios comes close to the observations, at least until 00:00 UTC 06 January 2016. However,
it was not the scenario with the highest probability, and, furthermore,
the spread in between the scenarios increased significantly with lead
time. Performing the ensemble runs and their analysis takes time.
The question to be answered here is if the forecast for days $2$
and $2.5$ can be improved if the ensemble members are selected by
their correlation with the observations in day one (c.f. Equation
\ref{eq:corr}). This is termed \textit{observation constrained ensemble forecast} from here onward. 
\begin{landscape}
\begin{figure}
\begin{centering}
\subfloat[$s_{thresh}=5$]{\protect\centering{}\protect\includegraphics[width=0.7\textwidth]{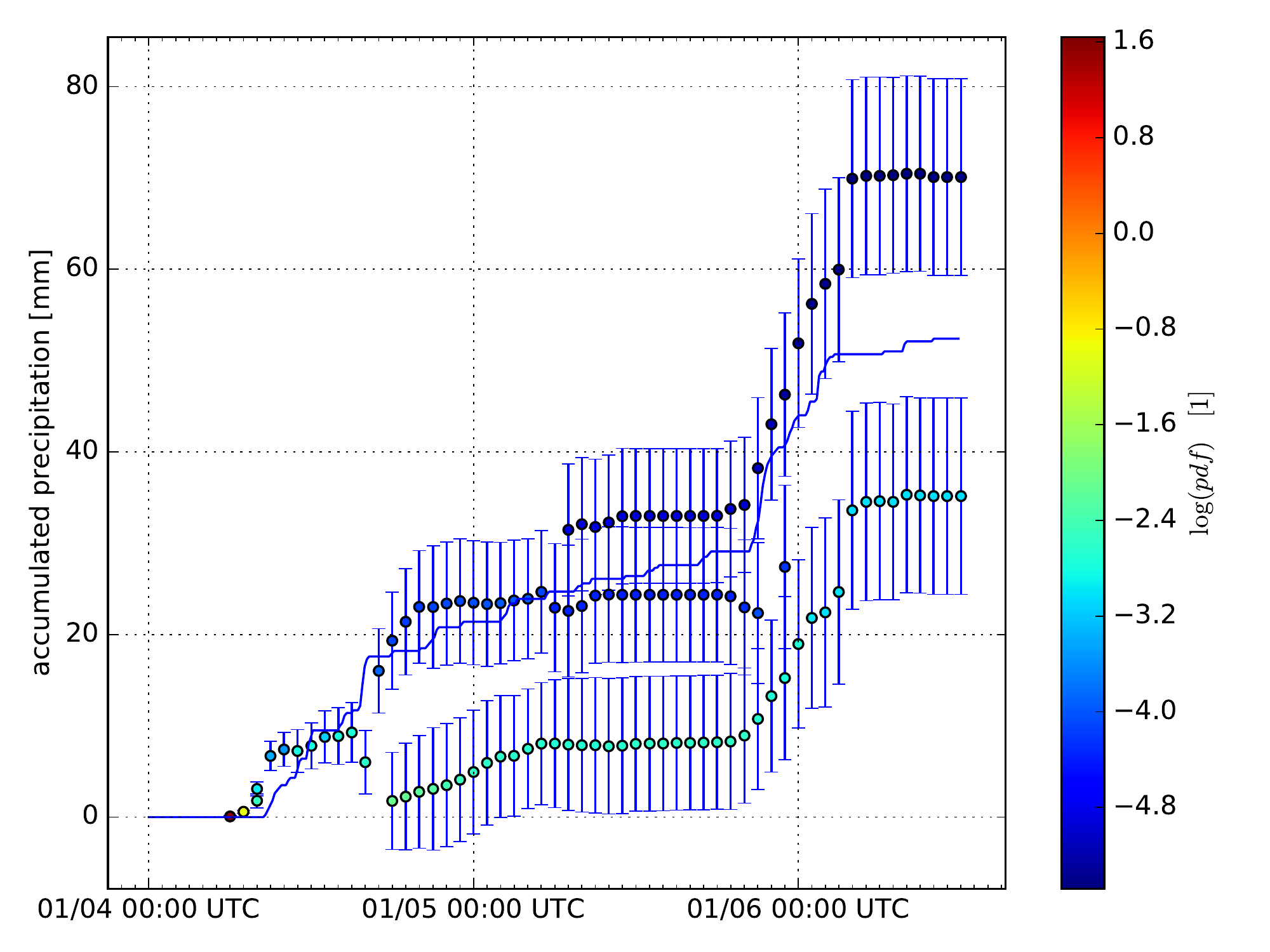}\protect}
\subfloat[$s_{thresh}=10$]{\protect\centering{}\protect\includegraphics[width=0.7\textwidth]{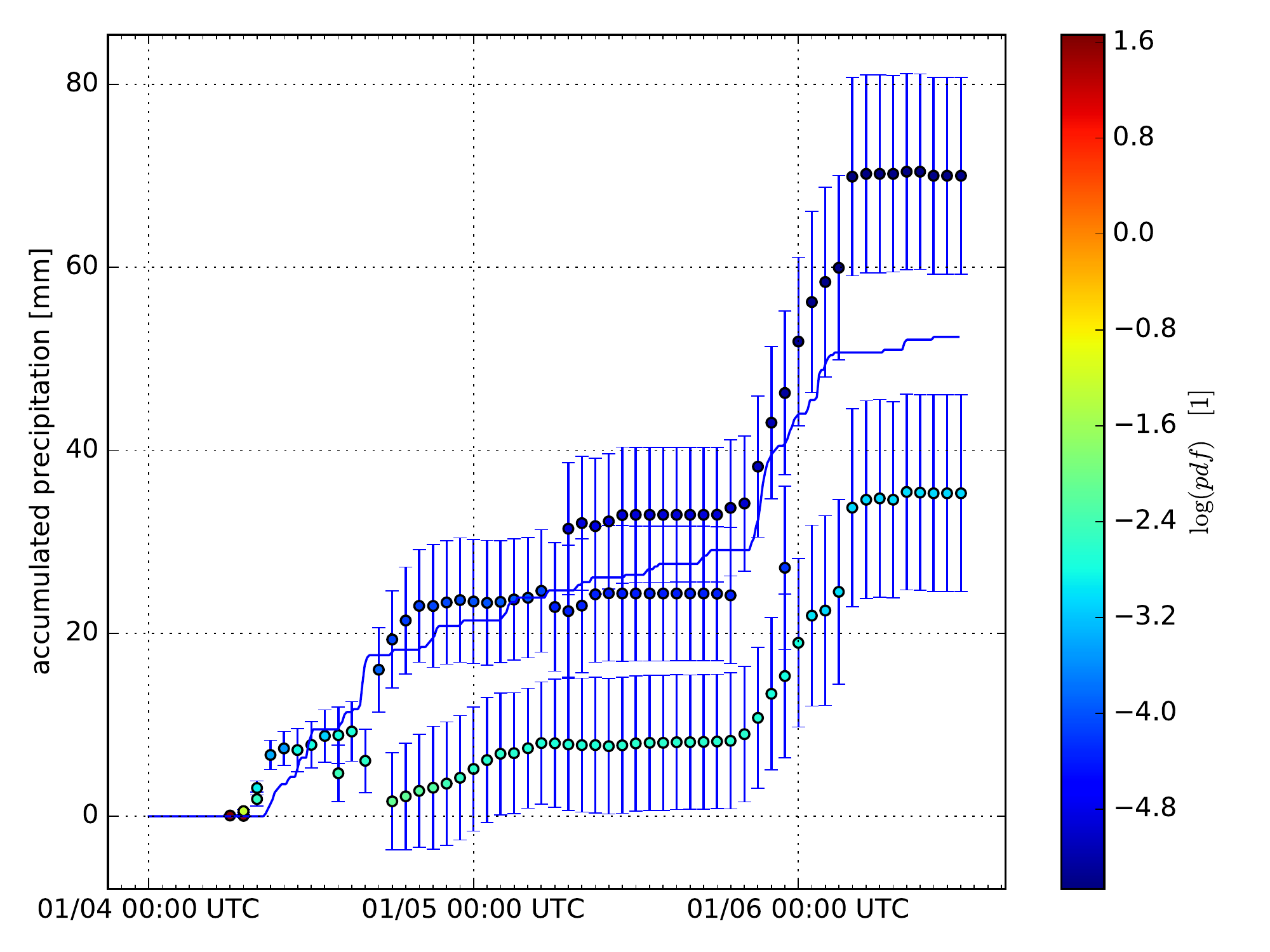}\protect}\\
\subfloat[$s_{thresh}=15$]{\protect\centering{}\protect\includegraphics[width=0.6\textwidth]{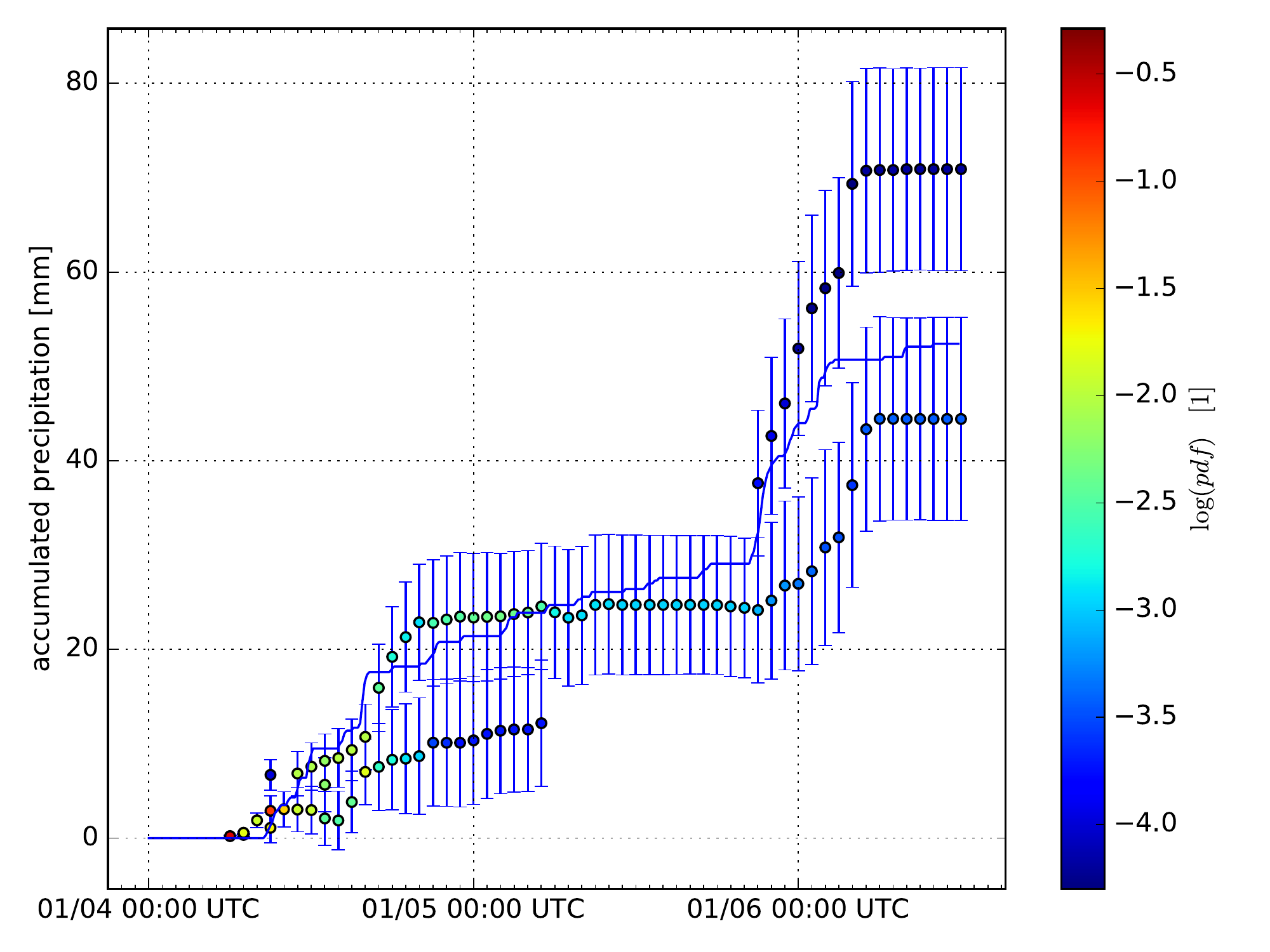}\protect}
\subfloat[$s_{thresh}=20$\label{fig:Selecting-ensemble-members-20}]{\protect\centering{}\protect\includegraphics[width=0.6\textwidth]{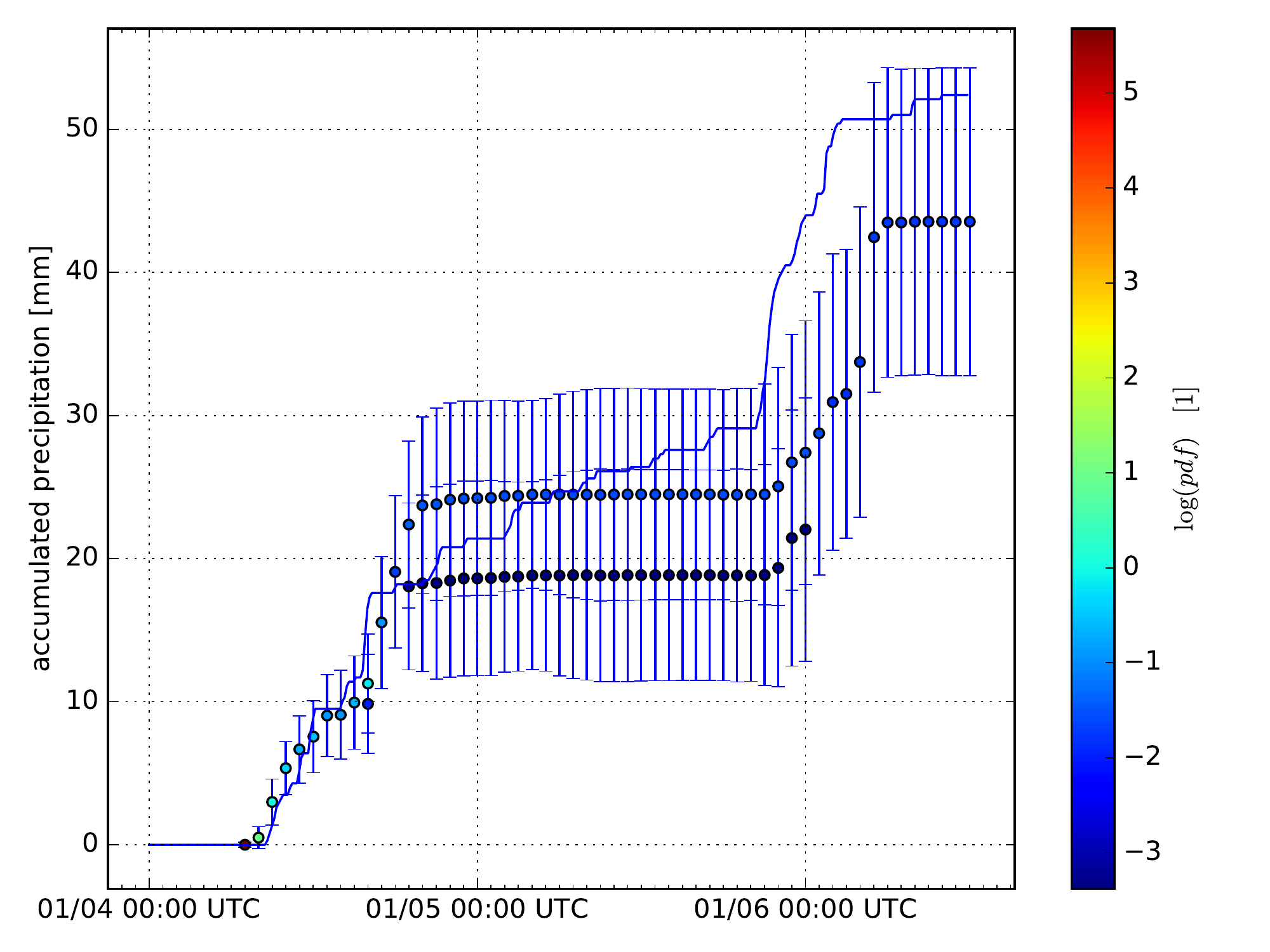}\protect}
\subfloat[$s_{thresh}=20$\label{fig:std_selected}]{\protect\centering{}\protect\includegraphics[width=0.6\textwidth]{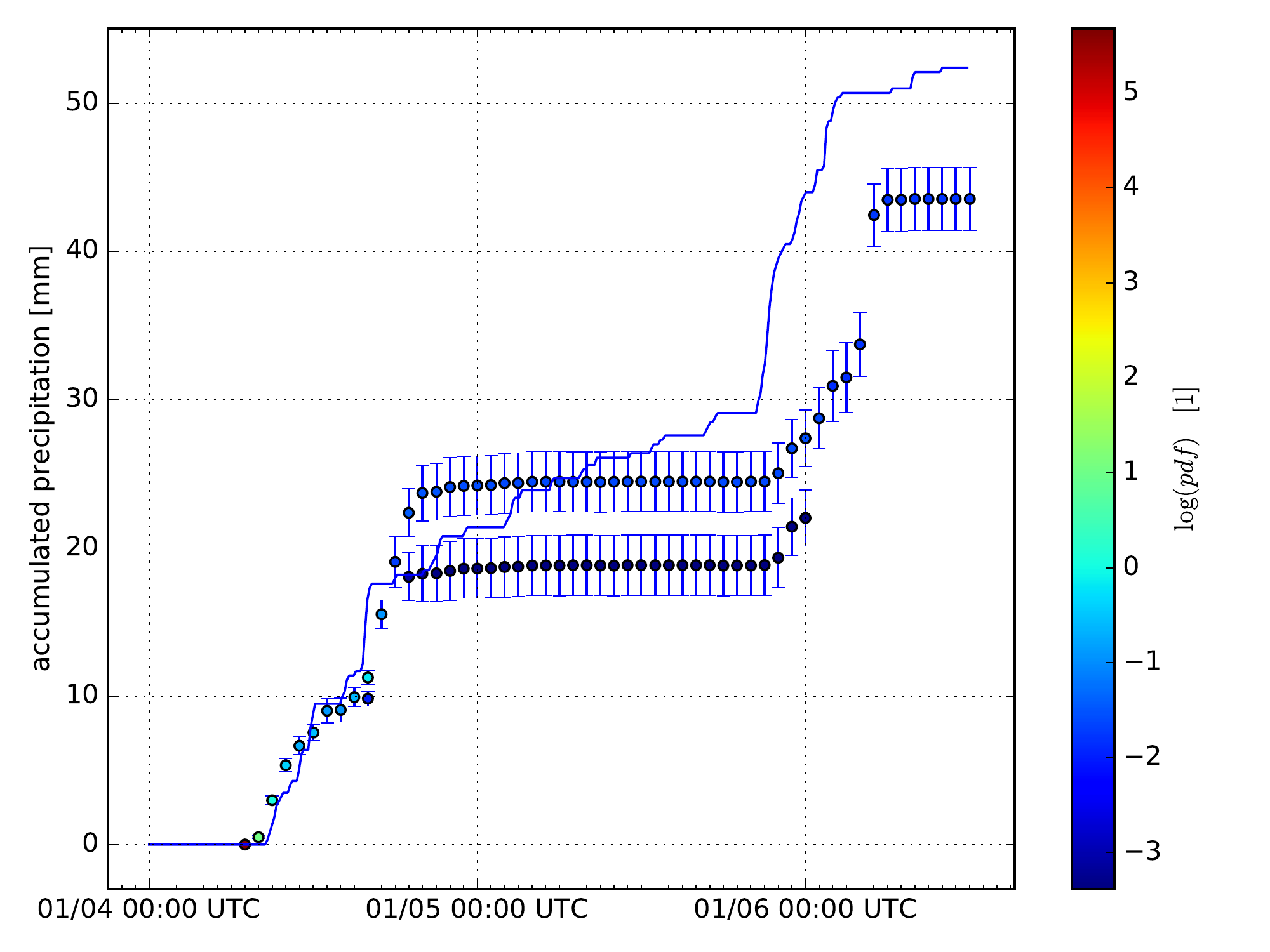}\protect}
\protect\caption{Selecting ensemble members according to the first 24 hrs\label{fig:Selecting-ensemble-members}.
The error bars in subpanel (a-d) represent one standard deviation of the original full
member ensemble. In subpanel e the error
bars denote the standard deviation of the selected ensemble members
only. }

\par\end{centering}
\end{figure}
\end{landscape}

\begin{figure}
\centerline{\includegraphics[width=0.7\textwidth]{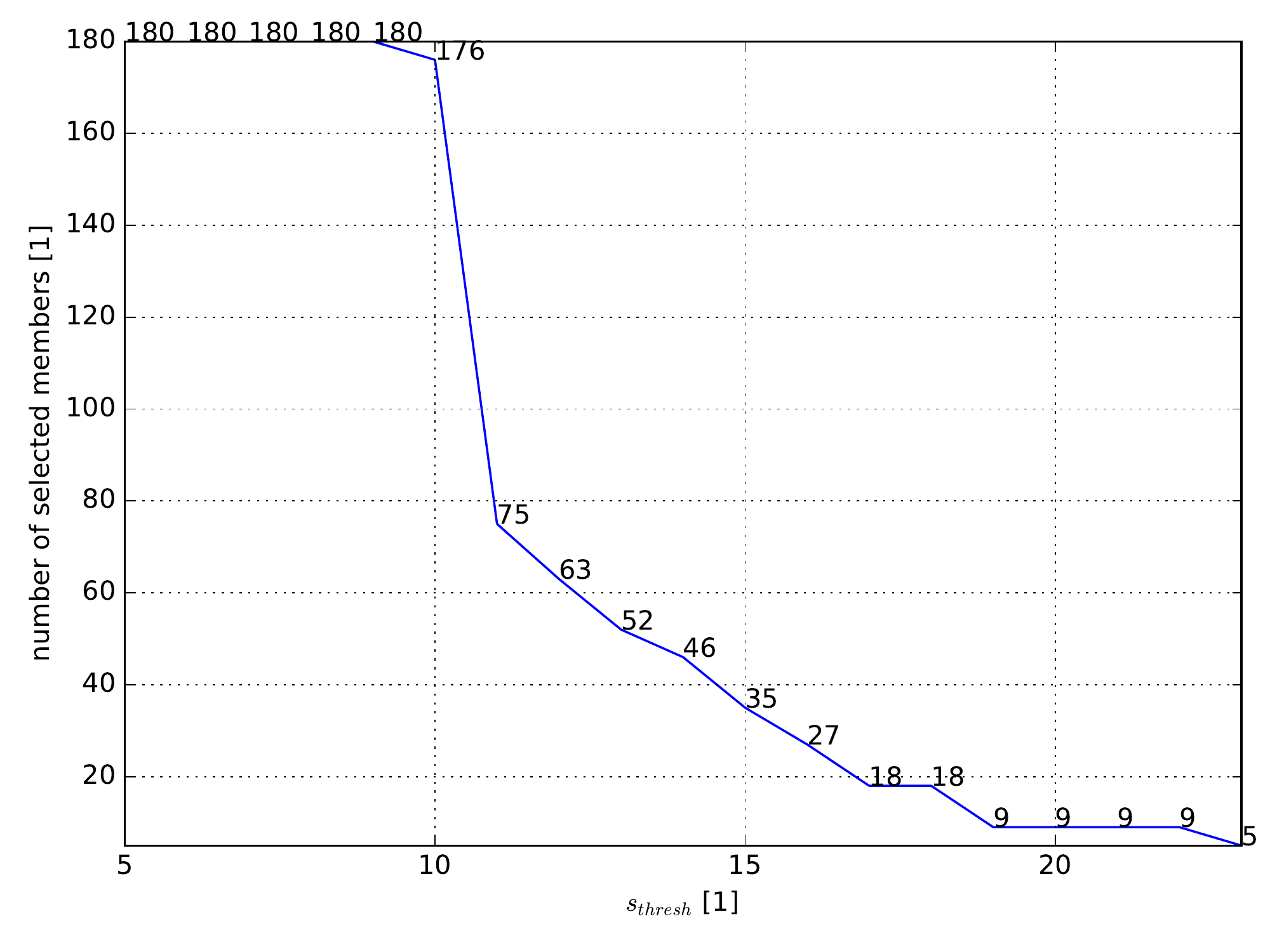}}
\caption{Number of ensemble members selected for the modal analysis above versus the minimum number of time steps within the first $24$hours that agree to within $\pm 1$ standard deviation with the observations, $s_{thresh}$, as in Equation~\ref{eq:corr}.\label{fig:remain}}
\end{figure}

Figure \ref{fig:Selecting-ensemble-members} shows how the
constraining of the ensemble yields an improved forecast.
With the requirement that at least $20$ time steps of the first $24$
be within two standard deviations the severely overestimating branch
has been removed and the observations are within less than one standard
deviation of the maxima. Figure \ref{fig:std_selected} reproduces
the plot for a match of more than $20$ time steps and plots the standard
deviation of the selected ensemble members as error bars. The number of ensemble members remaining after selection is shown in Figure~\ref{fig:remain}.

\section{Conclusions}

It was shown how a single deterministic forecast can be augmented
with probabilities and analyzed with respect to likely outcomes. At
small domains the effect of the lateral boundary condition dominates,
as expected. With larger domains the ensemble spread widens. For some
cases the \textit{ensemble enhanced mesoscale forecast }already provides
good guidance, as shown with the example of the 7 January 2016 case.
The 04 January 2016 case was more challenging. Extending the ensemble,
acknowledging a spatial uncertainty, slightly improved the guidance
in the \textit{subgrid ensemble enhanced mesoscale forecast.} However,
it still failed to emphasize that the high ($50$mm, in contrast
to the extreme $80$mm) precipitation event may be more justified.
Analyzing the local maxima in the cpd effectively highlighted these alternate
scenarios and differentiated them from the remaining ensemble members.
It would have been impossible to deduce this from the line plots of
the whole extended ensemble (Figure \ref{fig:Jan-4-geographic-lines}).
Finally it was shown how an \textit{observation constrained ensemble
forecast } significantly improved the longer lead time of the forecast
(fh$24$ to fh$60$), taking the observations available in the first
$24$hours into account.

It is hoped that these ideas will provide a product which will aid
the forecaster when deciding, from the mix of different models and sources
of information available, on balanced conclusion that lead to more 
accurate forecasts.

\section*{Acknowledgments}

The author would like to acknowledge the National Super Computer Center
(CNS) of IPCYT A.C. for providing access to the super computer facility
Thubat Kaal, and Dr Julio Scheinbaum of CICESE for kindly providing
access to the chaman cluster.
The author would like to thank Dr Magar and Dr Turrent Thompson for the constructive discussions and proofreading of the manuscript.

\end{document}